\documentclass[%
 aip,
 amsmath,amssymb,
 reprint,%
]{revtex4-1}
\usepackage{graphicx}
\usepackage{dcolumn}
\usepackage{bm}
\usepackage{color}
\usepackage[utf8]{inputenc}
\usepackage[T1]{fontenc}
\usepackage{mathptmx}
\usepackage{multirow}
\usepackage{float}
\usepackage{titlesec}
\titleclass{\subsubsubsection}{straight}[\subsection]
\newcounter{subsubsubsection}[subsubsection]
\renewcommand\thesubsubsubsection{\thesubsubsection.\alph{subsubsubsection}}
\titleformat{\subsubsubsection}
 {\normalfont\normalsize\bfseries}{\thesubsubsubsection}{1em}{}
\titlespacing*{\subsubsubsection}
{0pt}{3.25ex plus 1ex minus .2ex}{1.5ex plus .2ex}

\begin{document}
\preprint{AIP/123-QED}

\title{Two-fluid discrete Boltzmann model for compressible flows: based on Ellipsoidal Statistical Bhatnagar-Gross-Krook}

\author{Dejia Zhang}
 \affiliation{State Key Laboratory for GeoMechanics and Deep Underground Engineering, China University of Mining and Technology, Beijing 100083, P.R.China}
 \affiliation{Laboratory of Computational Physics, Institute of Applied Physics and Computational Mathematics, P. O. Box 8009-26, Beijing 100088, P.R.China}

\author{Aiguo Xu}%
 \thanks{Corresponding author: Xu\_Aiguo@iapcm.ac.cn}
\affiliation{Laboratory of Computational Physics, Institute of Applied Physics and Computational Mathematics, P. O. Box 8009-26, Beijing 100088, P.R.China}
\affiliation{Center for Applied Physics and Technology, MOE Key Center for High Energy Density Physics Simulations, College of Engineering, Peking University, Beijing 100871, P.R.China}
\affiliation{State Key Laboratory of Explosion Science and Technology, Beijing Institute of Technology, Beijing 100081, China}

\author{Yudong Zhang}
\affiliation{School of Mechanics and Safety Engineering, Zhengzhou University, Zhengzhou 450001, P.R.China}

\author{Yingjun Li}
 \thanks{Corresponding author: lyj@aphy.iphy.ac.cn}
\affiliation{State Key Laboratory for GeoMechanics and Deep Underground Engineering, China University of Mining and Technology, Beijing 100083, P.R.China}%

\date{\today}

\begin{abstract}
A two-fluid Discrete Boltzmann Model(DBM) for compressible flows based on Ellipsoidal Statistical Bhatnagar-Gross-Krook(ES-BGK) is presented. The model has flexible Prandtl number or specific heat ratio. Mathematically, the model is composed of two coupled Discrete Boltzmann Equations(DBE). Each DBE describes one component of the fluid. Physically, the model is equivalent to a macroscopic fluid model based on Navier-Stokes(NS) equations, and supplemented by a coarse-grained model for thermodynamic non-equilibrium behaviors.
To obtain a flexible Prandtl number, a  coefficient is introduced in the ellipsoidal statistical distribution function to control the viscosity. To obtain a flexible specific heat ratio, a parameter is introduced in the energy kinetic moments to control the extra degree of freedom. For binary mixture, the correspondence between the macroscopic fluid model and the DBM may be several-to-one.
  Five typical benchmark tests are used to verify and validate the model. Some interesting non-equilibrium results, which are not available in the NS model or the single-fluid DBM, are presented.
\end{abstract}

\maketitle

\section{\label{sec:level1} Introduction}

The complex non-equilibrium flows are common in the nature and engineering field. Numerical simulation has become an indispensable measure for complex flows\cite{2002Boltzmann,Xu2018-Chap2,A2017turbulence,Martin2012Numerical,2019Andreas-microchannel,2012Dommone,2008Gate,
Li2016Lattice,1997Gonnella,2020Lu-lattice,2006Kinetic,Li2015Rarefied,1994Bird-Molecular,liu2016-Boltzmann,
2015Chen-Boltzmann,2012Meng-Lattice}. Generally speaking, there are three kinds of physical models for flows: the macroscopic model, mesoscopic model and microscopic model.

The macroscopic models, based on Euler equations or Navier-Stoke(NS) equations, have long been applied to the large scale and slow behaviors in fluid mechanics \cite{2016Fluid-mechanics}. However, in some complex flows, due to shock waves or  detonation waves, large gradients of macroscopic quantities produces on both sides of the wavefront, and the interface regimes show strong Thermodynamic Non-Equilibrium (TNE) effects\cite{Xu2018-Chap2,Liu2016FP,Liu2017PRE}. Besides, compared with the interface width, the mean distance between neighboring fluid particles is not negligibly small, which challenges the physical rationality of continuity hypothesis which is the basestone of macroscopic models\cite{1965shock-wave-formation}. For example, in the Inertial Confined Fusion(ICF), there are many interactions between shock wave interfaces and material interfaces, which can not be measured accurately by macroscopic models\cite{2017ICF-instability}. In the aerospace field, the spacecraft would pass through the gas zones with different Knudsen numbers, which requires a model with cross-basin adaptive ability\cite{2012Rarefied-Gases,Celiberto2016PSST}. It has also been well-known that  microscale flows\cite{1992plasma-etching,2004Small-Devices,2012-fabrication} such as Micro-Electro-Mechanical System(MEMS)\cite{1998MEMS}, and oil flows in micropores often show a different flow and heat transfer characteristics, which beyond the description of macroscopic models.

In principle, the microscopic models, such as molecular dynamics\cite{Coninck2008Wetting,Cieplak2000Molecular,2010KHI}, are capable of capturing much more detailed information of the flows. Unfortunately, they are restricted to too small spatio-temporal scales due to the computing capability of the available computers. Consequently, the structures and dynamic behaviors of intermediate scales have long been remained a difficult problem. To investigate the behaviors of intermediate scales, a mesoscopic kinetic model is preferred. The recently proposed Discrete Boltzmann Model(DBM)\cite{Xu2018-Chap2,2018Gan-pre,2019Zhang-Shakhov} is one in such a category.

\emph{DBM is a coarse-grained modeling method. It selects a set of kinetic properties, described by kinetic moments of the distribution function $f$, to study the system. The set of kinetic properties compose a research perspective.}
In current DBM theory the Chapman-Enskog(CE) multiscale expansion\cite{1990Chapman} is one of the main ways to quickly determine the necessary kinetic properties to be preserved. According to the Chapman-Enskog analysis, via using higher order terms in Knudsen number, the DBM can be constructed for flows with higher degrees of TNE\cite{Xu2018-Chap2}.
As a mesoscopic model in physical description capability, a DBM may beyond the NS model from one or both the following two sides, (i) being applicable to deeper non-equilibrium flows, and/or (ii) bringing more kinetic information on the non-equilibrium flow. When a DBM adopts up to the second or higher order term in Knudsen number, it beyond the NS from both the two sides. If a DBM adopts only to the first order term in Knudsen number, it beyond the NS only from side (ii). In such a case, a DBM is equivalent to a NS model supplemented by a coarse-grained model for TNE behaviors. The NS model describes the conservative kinetic moments, i.e., the density, momentum, and energy in the evolution, while the coarse-grained model for TNE describes the evolution of corresponding nonconservative kinetic moments. The latter are used to supplement the shortage of the former in capturing non-equilibrium behaviors.

The most fundamental step to starting the physical function, side (ii), is to use the nonconserved kinetic moments of $(f - f ^{eq})$ to describe the specific deviation from thermodynamic equilibrium state of the system behavior, which was suggested by Xu, et al. in 2012\cite{Xu2012-FoP-review}, where $f^{eq}$ is the corresponding equilibrium distribution function. Then, it was suggested to investigate the complex TNE behaviors in the phase space opened by the independent components of the nonconserved kinetic moments of $(f - f ^{eq})$ and its subspaces\cite{2015Xu-PRE}. In the phase space opened by nonconserved kinetic moments and its subspaces,  corresponding non-equilibrium strength was defined by means of the distance from the origin, and non-equilibrium state similarity and kinetic process similarity were defined by means of the reciprocal of the distance between two points
\cite{2015Xu-PRE,Xu2018-RGD31}. Via those concepts some previously unextractable information can be hierarchical, quantitative research.

The DBM has been applied to many complicated fluid systems, such as fluid instability\cite{2019Zhang-Shakhov,2016Lai-DBMRT,2016Chen-RTI,2017Lin-DDBM-RT,2019Gan-KHI,2020Ye-RTI}, compressible flow under impact\cite{Xu2018-Chap2,2018Gan-pre,2019Zhang-Shakhov,2018Chen-PoF}, non-equilibrium combustion\cite{2015Xu-PRE,2016Lin-CNF,2016Zhang-CNF}, multi-phase flow and non-equilibrium phase transition\cite{2015Gan-Soft,2019Zhang-Soft} and brought a series of new insights in related fields. Besides by theoretical analyses and experimental data \cite{Lin2017SR}, some of the DBM results have been confirmed and supplemented by simulation results of molecular dynamics \cite{Liu2016FP,Liu2017PRE}, and direct simulation Monte Carlo\cite{2019Zhang-Shakhov,Sebastiao2018CNF,Gimelshein2019PRF}, etc.

Roughly speaking, according to the physical identification capability, there are two kinds of fluid models: single-fluid and multi-fluid model. The single-fluid macroscopic model uses a set of hydrodynamic quantities, (density $\rho$, flow velocity $\bf{u}$, temperature $T$, pressure $p$), to describe the system. It ignores the difference of components and regards that the fluid system consists only of a single-component. It is the simplest fluid model.
The $N$-fluid macroscopic model uses $N$ set of hydrodynamic quantities, (density $\rho^{\sigma}$, flow velocity $\bf{u^{\sigma}}$, temperature $T^{\sigma}$, pressure $p^{\sigma}$), to describe the system, where $\sigma$ is the index of the fluid component. Consequently, compared with single-fluid model, two-fluid model is a finer description and can simulate more precisely the fluid system which is composed of two different components. For example, Fan et al. proposed an ion-electron non-equilibrium model, indicating the existence of ion-electron non-equilibrium in the hot spot of high-foot implosions, which can not be obtained from single-fluid model\cite{2016Fan,2017Fan}. Currently, many works have been done in multi-fluid model\cite{Xu2005PRE,Lin2018Binary,Arcidiacono2007PRE,Liu2016JCP} and have made significant progress in multi-phase flows\cite{Fei2019POF,Bertevas2019POF}, fluid instability\cite{2017Lin-DDBM-RT,2019Kinetic}, reactive flows\cite{Chen2015IJHMT,Hosseini2018PA,Lin2017SR}, and combustion \cite{2016Lin-CNF,2018Mesoscopic}.
Correspondingly, the single-fluid DBM uses a single distribution function to describe the system\cite{2016Lai-DBMRT,2016Chen-RTI}. The $N$-fluid DBM uses $N$ distribution functions to describe the system. Each distribution function describes one fluid component\cite{2017Lin-DDBM-RT,2016Lin-CNF}. Currently, the two-fluid DBM have made significant progress in combustion, fluid instability and other non-equilibrium flows. \cite{2016Lin-CNF,2017Lin-DDBM-RT,2019Kinetic,Lin2018Binary,2018Mesoscopic,Lin2017SR}.

It is known that the Prandtl number in simplified Boltzmann equation based on the Bhatnagar-Gross-Krook (BGK) model \cite{BGK1954} is fixed to unity. As a result, in the model system based on the BGK, the viscosity and heat conductivity change simultaneously when the relaxation time is adjusted\cite{2016Lin-CNF}. To remove this binding between viscosity and heat conductivity, there are two solutions. The first solution is to construct Multiple-Relaxation-Time(MRT) collision model\cite{2016Chen-RTI,2015Xu-PRE}. The second is to keep the single-relaxation-time framework and introduce a parameter in the collision term to control the viscosity and/or heat conductivity\cite{2019Zhang-Shakhov,1966ES,2017Zhang-ES,1968Shakhov,1990Liu}. To ensure the relaxation times have clear physical correspondences, the MRT model is generally first calculated in the kinetic moment space and then transformed back to the discrete velocity space.
It should be pointed out that the models \cite{2019Zhang-Shakhov,1966ES,2017Zhang-ES,1968Shakhov,1990Liu} in the second solution for a flexible Prandtl number are all single-fluid models. It is meaningful to develop them to two--fluid models.

In this work, we develop a two-fluid DBM based on the Ellipsoidal Statistical BGK (ES-BGK) model\cite{1966ES}, which is an extension of the single-fluid DBM proposed by Zhang, et al. \cite{2017Zhang-ES}. The paper is organized as follows: Section \ref{Model construction} presents the model construction. Section \ref{Numerical simulations} verifies and validates the new model. Section \ref{Conclusions} concludes the paper.

\section{Model construction}\label{Model construction}
Based on the ES-BGK single-relaxation model, we present a two-fluid DBM for compressible flows with a flexible Prandtl number and specific heat ratio. To construct a two-fluid DBM from Boltzmann equation, three steps are needed. The first step is to simplify the collision operator. The most common practice is the collision operator linearization. Values of the least amount of kinetic moments, which are necessary for describing the flow system, of collision operator must remain unchanged for the integral-form and the linearized-form cases in this simplifying process. The kinetic moments we need are necessary to rely on the specific physical problem which under consideration.  Generally speaking, the deeper the non-equilibrium flows, the more complex the flow behaviors, and the more kinetic moments are necessary. In any non-equilibrium flows, the initial several kinetic moments are necessary, including the three conserved kinetic moments (density, momentum and energy).

For binary mixture, there are two kinds of collision models, the one-step (relaxation collision) model and two-step (relaxation collision) model. The basic assumption of the two-step (relaxation collision) model is that each component first experiences equilibration, then the mixture experiences equilibration. When the particle masses of the two components are different, the temporal evolution of the binary mixture is described by the formal two-step model as occurring in three epochs. Firstly, the component with lighter particle mass experiences equilibration, then the component with heavier particle mass experiences equilibration, and finally the whole system experiences equilibration\cite{Xu2005PRE}.

For the two-fluid kinetic model, in this work, we start from the following ES-BGK Boltzmann equation with one-step (relaxation collision) model,
\begin{equation}
\frac{\partial f^{\sigma}}{\partial t}+\mathbf{v}\cdot\frac{\partial f^{\sigma}}{\partial \mathbf{r}}=-\frac{1}{\tau^{\sigma}}(f^{\sigma}-f^{\sigma,ES})
\tt{.}
\end{equation}
where $\sigma=A \;or \; B$ is the index of the component; $\mathbf{v}$, $\mathbf{r}$ and $\mathbf{a}$ represent velocity vector, space vector and acceleration vector, respectively; $\tau^{\sigma}$ is the relaxation time.
The distribution function at the point ($\mathbf{r}$,$\mathbf{v}$) in phase space reads $f^{\sigma}(\mathbf{r},\mathbf{v})$.
$f^{\sigma,ES}$ gives the evolving direction of $f^{\sigma}$, it takes the continuous form as follow:
\begin{equation}
\begin{aligned}
f^{\sigma,ES}&=n^{\sigma}(\frac{m^{\sigma}}{2\pi})^{\frac{D}{2}}\frac{1}{\sqrt{\left|\lambda_{\alpha\beta}\right|}}(\frac{m^{\sigma}}{2\pi I^{\sigma}T})^\frac{1}{2}\\
&\times \exp[-\frac{m^{\sigma}(\mathbf{v}-\mathbf{u})^2}{2\left|\lambda_{\alpha\beta}\right|}-\frac{m^{\sigma}\eta^{\sigma2}}{2I^{\sigma}T}]\tt{,}
\end{aligned}
\end{equation}
where $D$ is the spatial dimension. The quantity $n^{\sigma}$, $m^{\sigma}$, $T$ and $\mathbf{u}$ represent the particle number density of $\sigma$, particle mass of $\sigma$, temperature of the physical system (the mixture), and velocity vector of the physical system, respectively. $I^{\sigma}$ and $\eta^{\sigma}$ represent the extra degree of freedom and extra energy of freedom, respectively. The modified term $\lambda_{\alpha\beta}=kT\delta_{\alpha\beta}+\frac{b^{\sigma}}{n^{\sigma}}\Delta^{\sigma*}_{2,\alpha\beta}$, where $k$ is the Boltzmann constant and $\Delta^{\sigma*}_{2,\alpha\beta}$ represents viscous stress. $b^{\sigma}$ is an adjustable coefficient. The ES distribution $f^{\sigma,ES}$ is equal to Maxwellian distribution $f^{\sigma,eq}$ when $b^{\sigma}=0$. Thus, the Prandtl number and specific heat ratio are flexible by adjusting coefficient $b^{\sigma}$ and parameter $I^{\sigma}$, It should be noticed that when adjusting one of them($b^{\sigma}$ and $I^{\sigma}$), the other must be fixed to zero(as shown in appendix \ref{App}).
Based on the same one-step model, the formulated DBM will be unique only if the necessary kinetic moment relations are fixed. However, we will show that the hydrodynamic equations obtained from the Chapman-Enskog analysis may be in different forms. The hydrodynamic equations with different forms correspond to the same DBM.

For the mixture, there are two kinds flow velocities: flow velocity of component ${\sigma}$ denoted by $\mathbf{u^{\sigma}}$ and flow velocity of mixture denoted by $\mathbf{u}$. The particle number density, particle mass, and velocity of component $\sigma$ are defined as
\begin{equation}
n^{\sigma}=\sum_{i}f^{\sigma}_{i}\tt{,}\rho^{\sigma}=n^{\sigma}m^{\sigma} \tt{,}
\end{equation}
\begin{equation}
\mathbf{u}^{\sigma}=\frac{\sum_{i}f^{\sigma}_{i}\mathbf{v}_{i}}{n^{\sigma}} \tt{,}
\end{equation}
The particle number density, particle mass, and velocity of mixture are defined as follows:
\begin{equation}
n=\sum_{\sigma}n^{\sigma},\rho=\sum_{\sigma}\rho^{\sigma} \tt{,}
\end{equation}
\begin{equation}
\mathbf{u}=\frac{\sum_{\sigma}\rho^{\sigma}\mathbf{u}^{\sigma}}{\rho}
\end{equation}

Because the definition of internal energy (temperature) depends on the flow velocity chosen as a reference, we can define the internal energy (temperature) in two different ways. The first definition is $E^{\sigma*}_{I}=\frac{1}{2}m^{\sigma}\sum\limits_{i}f^{\sigma}_{i}((\mathbf{v}_{i}-\mathbf{u})^2+\eta^{2}_{i})$. The corresponding definition of temperature for component $\sigma$ and the mixture are as follows:
\begin{equation}
T^{\sigma*}=\frac{2E_{I}^{\sigma*}}{n^{\sigma}(D+I^{\sigma})} \tt{,}
\end{equation}
\begin{equation}
T=\frac{2E_{I}^{*}}{\sum_{\sigma}n^{\sigma} (D+I^{\sigma})}
\end{equation}
where the kinetic energy is $E_{K}^{\sigma*}=\frac{1}{2}\rho^{\sigma}\mathbf{u}\cdot\mathbf{u}$ and $E_{I}^{*}=\sum_{\sigma}E_{I}^{\sigma*}$.
In this work we focus on the case of ideal gas. Thus,
the definition of pressure for  component $\sigma$ and the mixture are $p^{\sigma*}=n^{\sigma}T^{\sigma*}$ and $p^{*}=\sum_{\sigma}n^{\sigma}T^{\sigma *}=nT$, respectively. We can also define the internal energy as $E^{\sigma}_{I}=\frac{1}{2}m^{\sigma}\sum\limits_{i}f^{\sigma}_{i}((\mathbf{v}_{i}-\mathbf{u}^{\sigma})^2+\eta^{2}_{i})$. The corresponding definition of temperature for component $\sigma$ and mixture are
\begin{equation}
T^{\sigma}=\frac{2E_{I}^{\sigma}}{n^{\sigma}(D+I^{\sigma})} \tt{,}
\end{equation}
\begin{equation}
T=\frac{2(E_{I}+\Delta E_{I}^{*})}{\sum_{\sigma}n^{\sigma} (D+I^{\sigma})}
\end{equation}
where
\begin{equation}
\Delta E_{I}^{*}=E_{I}^{*}-E_{I}=\frac{\rho^{A}\rho^{B}(u^{A}_{\alpha}-u^{B}_{\alpha})^{2}}{2(\rho^{A}+\rho^{B})} \tt{,}
\end{equation}
is the difference between $E_{I}^{*}$ and $E_{I}$
\footnote[65]{If we transfer from the second definition of internal energy to the first definition, the amount of energy, $\Delta E_{I}^{*}$, will be transformed from kinetic energy to internal energy.}.
In this way, the kinetic energy is $E_{K}^{\sigma}=\frac{1}{2}\rho^{\sigma}\mathbf{u}^{\sigma}\cdot\mathbf{u}^{\sigma}$. The pressure definition of  component $\sigma$ and mixture are $p^{\sigma}=n^{\sigma}T^{\sigma}$ and $p=nT$, respectively. As we can see, when the velocities of two components approach the same, then $T^{\sigma*}=T^{\sigma}=T$.
The first definition can be seen in many works\cite{2016Lin-CNF,Xu2005PRE}. We choose the first definition in our paper.

The second step is to discretize the velocity space, then we can get the discrete ES-Boltzmann-BGK equation:
\begin{equation}
\frac{\partial f^{\sigma}_{i}}{\partial t}+v_{i\alpha}\cdot\frac{\partial f^{\sigma}_{i}}{\partial r_{i\alpha}}=-\frac{1}{\tau^{\sigma}}(f^{\sigma}_{i}-f^{\sigma,ES}_{i})\tt{,}
\end{equation}
where $f^{\sigma}_{i}(\mathbf{r},\mathbf{v})$ is the discrete distribution function with $i=1$, $2$, $\cdots$, $N$ and $N$ is the total number of the discrete velocities. On the condition of remaining values of some specific kinetic moments unchanged, we can substitute the velocity space by a limited number of particle velocities according the discrete Boltzmann method. The specific kinetic moments that needs to be satisfied depend on the specific physical problems. As an initial step, in this work, we develop a two-fluid DBM where only the first order thermodynamic non-equilibrium effects are taken into account. In this case, only the $0^{th}$ order to $(4,2)^{th}$ order kinetic moments are necessary according to the CE analysis, where ``4,2'' means that the $4^{th}$ order tensor is contracted to a $2^{nd}$ order tensor.  Similar subscripts ``3,1'' will also be used in the following part of the paper.

The third step in constructing a DBM is to present a solution for describing non-equilibrium state and extracting non-equilibrium information. Besides recovering the NS model, a DBM can describe TNE behaviors which are not available in an NS model. The most fundamental TNE information can be extracted from the nonconserved kinetic moments of $(f^{\sigma} - f^{\sigma,eq})$, based on which various characteristic quantities can be defined for describing the TNE state from different perspectives. We first define the following TNE quantities,
\begin{equation}
\bm{\Delta}^{\sigma *}_{2}=m^{\sigma}\sum_{i}(f^{\sigma}_{i}-f^{\sigma,eq}_{i})\mathbf{v}^{*}_{i}\mathbf{v}^{*}_{i} \tt{,} \label{Eq:DDBM-NF28}
\end{equation}
\begin{equation}
\bm{\Delta}^{\sigma *}_{3,1}=\frac{1}{2}m^{\sigma}\sum_{i}(f^{\sigma}_{i}-f^{\sigma,eq}_{i})(\mathbf{v}^{*}_{i}\cdot\mathbf{v}^{*}_{i}+\eta_{i}^{\sigma 2})\mathbf{v}^{*}_{i} \tt{,}
\end{equation}
\begin{equation}
\bm{\Delta}^{\sigma *}_{3}=m^{\sigma}\sum_{i}(f^{\sigma}_{i}-f^{\sigma,eq}_{i})\mathbf{v}^{*}_{i}\mathbf{v}^{*}_{i}\mathbf{v}^{*}_{i} \tt{,}
\end{equation}
\begin{equation}
\bm{\Delta}^{\sigma *}_{4,2}=\frac{1}{2}m^{\sigma}\sum_{i}(f^{\sigma}_{i}-f^{\sigma,eq}_{i})(\mathbf{v}^{*}_{i}\cdot\mathbf{v}^{*}_{i}+\eta_{i}^{\sigma 2})\mathbf{v}^{*}_{i}\mathbf{v}^{*}_{i} \tt{,}
\end{equation}
$\mathbf{v}^{*}_{i}=\mathbf{v}_{i}-\mathbf{u}$ denotes the central velocity, where $\mathbf{u}$ represents the macro flow speed of system. The first subscript of $\bm{\Delta}^{\sigma*}$ represents the number of velocity  $\mathbf{v}^{*}_{i}$ and the second is the order of tensor. Physically, the tensors $\bm{\Delta^{\sigma *}_{2}}=\Delta^{\sigma *}_{2,\alpha\beta}\mathbf{e}_{\alpha}\mathbf{e}_{\beta}$ and $\bm{\Delta}^{\sigma *}_{3,1}=\Delta^{\sigma *}_{3,1}\mathbf{e}_{\alpha}$ represent viscous stress tensor and heat flux tensor, respectively, with $\mathbf{e}_{\alpha}$ the unit vector in the $\alpha$ direction. $\bm{\Delta}^{\sigma *}_{3}=\Delta^{\sigma *}_{3\alpha\beta\gamma}\mathbf{e}_{\alpha}\mathbf{e}_{\beta}\mathbf{e}_{\gamma}$ and $\bm{\Delta}^{\sigma *}_{4,2}=\Delta^{\sigma *}_{4,2\alpha\beta}\mathbf{e}_{\alpha}\mathbf{e}_{\beta}$
represent the flux of viscous stress and flux of heat flux, respectively, which
are higher order non-equilibrium quantities beyond traditional NS model.
Based on the most fundamental TNE quantities, in this work,
 we further introduce the following five condensed measures: $\left|\bm{\Delta}_{2}^{\sigma*}\right|$, $\left|\bm{\Delta}_{3,1}^{\sigma*}\right|$, $\left|\bm{\Delta}_{3}^{\sigma*}\right|$, $\left|\bm{\Delta}_{4,2}^{\sigma*}\right|$, and $\overline{D}$. $\left|\bm{\Delta}_{2}^{\sigma*}\right|$ is used to measure the strength of the viscous stress; $\left|\bm{\Delta}_{3,1}^{\sigma*}\right|$ indicates the intensity of the heat flux; $\left|\bm{\Delta}_{3}^{\sigma*}\right|$ and $\left|\bm{\Delta}_{4,2}^{\sigma*}\right|$ represent the intensities of $\bm{\Delta}_{3}^{\sigma*}$ and $\bm{\Delta}_{4,2}^{\sigma*}$, respectively; $\overline{D}^{*}$ indicates the global average Thermodynamic Non-Equilibrium intensity, i.e., ``TNE'' strength, specifically,
\begin{equation}
\left|\bm{\Delta}_{2}^{\sigma*}\right|=\sqrt{\Delta_{2,xx}^{\sigma*2}+2\Delta_{2,xy}^{\sigma*2}+\Delta_{2,yy}^{\sigma*2}}
 \tt{,}
\end{equation}
\begin{equation}
\left|\bm{\Delta}_{3,1}^{\sigma*}\right|=\sqrt{\Delta_{3,1,x}^{\sigma*2}+\Delta_{3,1,y}^{\sigma*2}}
 \tt{,}
\end{equation}
\begin{equation}
\left|\bm{\Delta}_{3}^{\sigma*}\right|=\sqrt{\Delta_{3,xxx}^{\sigma*2}+3\Delta_{3,xxy}^{\sigma*2}
+3\Delta_{3,xyy}^{\sigma*2}+\Delta_{3,yyy}^{\sigma*2}}
 \tt{,}
\end{equation}
\begin{equation}
\left|\bm{\Delta}_{4,2}^{\sigma*}\right|=\sqrt{\Delta_{4,2,xx}^{\sigma*2}+2\Delta_{4,2,xy}^{\sigma*2}
+\Delta_{4,2,yy}^{\sigma*2}}
 \tt{,}
\end{equation}
\begin{equation}
\overline{D}^{\sigma *}=\sqrt{{\left|\bm{\Delta}_{2}^{\sigma*}\right|^2+\left|\bm{\Delta}_{3,1}^{\sigma*}\right|^2
+\left|\bm{\Delta}_{3}^{\sigma*}\right|^2+\left|\bm{\Delta}_{4,2}^{\sigma*}\right|^2}}
 \tt{,}
\end{equation}
More TNE quantities can be defined according to the need in practical applications of DBM\cite{Xu2018-Chap2}.

\subsection{The discrete form of $\mathbf{f}^{\sigma,ES}$}
For the convenience of simulation, the discrete form of $\mathbf{f}^{\sigma,ES}$ should be specified. Based on statistical mechanics, some kinetic moments of discrete distribution function can be written as follows:
\begin{equation}
M^{ES}_{0}=\sum_{i}f^{\sigma,ES}_{i}=n^{\sigma} \tt{,}
\end{equation}
\begin{equation}
M^{ES}_{1,\alpha}=\sum_{i}f^{\sigma,ES}_{i}v_{i\alpha}=n^{\sigma}u_{\alpha} \tt{,}
\end{equation}
\begin{eqnarray}
\begin{aligned}
M^{ES}_{2,0}=\sum_{i}f^{\sigma,ES}_{i}(v_{i\alpha}\cdot v_{i\alpha}&+\eta^{\sigma2}_{i})=\frac{n^{\sigma}}{m^{\sigma}}[\lambda_{\alpha\alpha}\\
&+m^{\sigma}u_{\alpha}\cdot u_{\alpha}]+n^{\sigma}I^{\sigma}\frac{T}{m^\sigma}   \tt{,} \label{Eq:DDBM-NF24}
\end{aligned}
\end{eqnarray}
\begin{equation}
M^{ES}_{2,\alpha\beta}=\sum_{i}f^{\sigma,ES}_{i}v_{i\alpha}v_{i\beta}=n^{\sigma}[\frac{\lambda_{\alpha\beta}}{m^{\sigma}}+u_{\alpha}u_{\beta}] \tt{,} \label{Eq:DDBM-NF25}
\end{equation}
\begin{eqnarray}
\begin{aligned}
M^{ES}_{3,1,\alpha}&=\sum_{i}f^{\sigma,ES}_{i}( v_{i\gamma}\cdot v_{i\gamma}+\eta^{2}_{i} ) v_{i\alpha}\\
&=M^{ES}_{3,\alpha\gamma\gamma}+n^{\sigma}I^{\sigma}u_{\alpha}\frac{T}{m^\sigma}
\tt{,} \label{Eq:DDBM-NF26}
\end{aligned}
\end{eqnarray}
\begin{eqnarray}
\begin{aligned}
M_{3,\alpha\beta\chi}^{ES}&=\sum_{i}f^{\sigma,ES}_{i}v_{i\alpha}v_{i\beta}v_{i\gamma}=n^{\sigma}u_{\alpha}u_{\beta}u_{\gamma}\\
&+\frac{n^{\sigma}}{m^{\sigma}}(u_{\alpha}\lambda_{\beta\gamma}+u_{\beta}\lambda_{\alpha\gamma}+u_{\gamma}\lambda_{\alpha\beta})
\end{aligned}
\end{eqnarray}
\begin{equation}
\begin{aligned}
M^{ES}_{4,2,\alpha\beta}&=\sum_{i}f^{\sigma,ES}_{i}( v_{i\gamma}\cdot v_{i\gamma}+\eta^{2}_{i})v_{i\alpha}v_{i\beta}=M^{ES}_{4,\alpha\beta\gamma\gamma}\\
&+I^{\sigma}\frac{T}{m^{\sigma2}}n^{\sigma}(\lambda_{\alpha\beta}+m^{\sigma}u_{\alpha}u_{\beta})\tt{.}
\end{aligned}
\end{equation}

where the second order tensor $M^{ES}_{4,2,\alpha\beta}$ is contracted from the fourth order tensor $M^{ES}_{4,\alpha\beta\gamma\chi}$ which reads
\begin{equation}
\begin{aligned}
M^{ES}_{4,\alpha\beta\gamma\chi}&=\sum_{i}f^{\sigma,ES}_{i}v_{i\alpha}v_{i\beta}v_{i\gamma}v_{i\chi}
=\frac{n^{\sigma}}{m^{\sigma2}}(\lambda_{\alpha\beta}\lambda_{\gamma\chi}+\lambda_{\alpha\gamma}\lambda_{\beta\chi}\\
&+\lambda_{\alpha\chi}\lambda_{\beta\gamma}+u_{\alpha}u_{\beta}\lambda_{\gamma\chi}+u_{\alpha}u_{\chi}\lambda_{\beta\gamma}+u_{\alpha}u_{\gamma}\lambda_{\beta\chi}\\
&+u_{\beta}u_{\chi}\lambda_{\alpha\gamma}+u_{\beta}u_{\gamma}\lambda_{\alpha\chi}+u_{\gamma}u_{\chi}\lambda_{\alpha\beta}\\
&+m^{\sigma2}u_{\alpha}u_{\beta}u_{\gamma}u_{\chi})
\tt{.}
\end{aligned}
\end{equation}

\begin{figure}[tbp]
\center\includegraphics*
[width=0.5\textwidth]{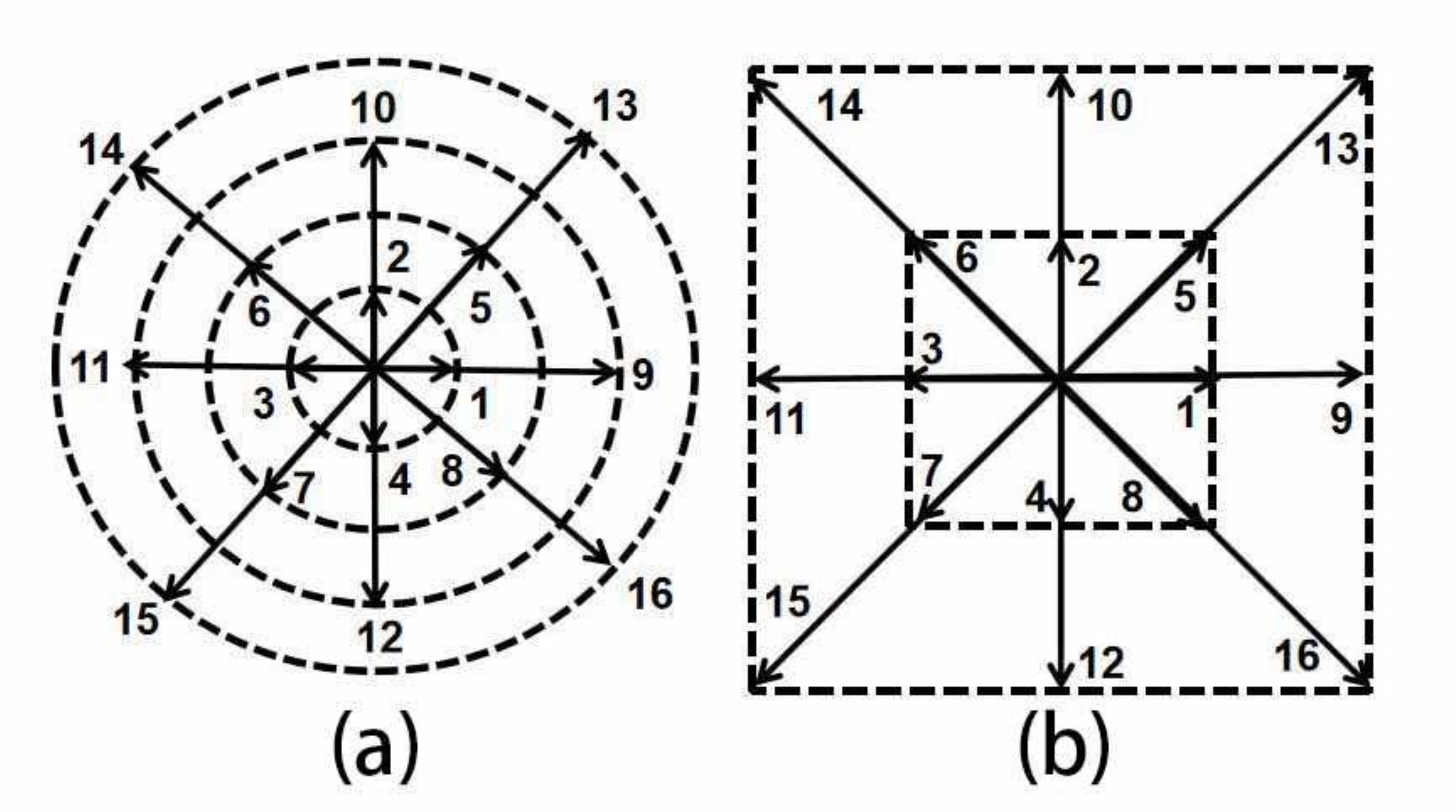}
\caption{Sketches (a) and (b) of two kinds D2V16 model used in the present paper, respectively. The numbers in the figure are the indexes of the discrete velocities.} \label{fig1}
\end{figure}

It should be noticed that Einstein summation convention is used and $b^{A}=b^{B}=b$ in the above equations. Actually, those kinetic moment equations can be written in a matrix form, i.e.,
\begin{equation}
\mathbf{C}\cdot\mathbf{f}^{\sigma,ES}=\mathbf{\hat{f}}^{\sigma,ES}
\tt{,}
\end{equation}
where $\mathbf{f}^{\sigma,ES}$ is a vector of discrete distribution function in velocity space and $\mathbf{\hat{f}}^{\sigma,ES}$ is the discrete distribution function in moment space. $\mathbf{v}_{i}$  represents discrete velocity. $\mathbf{C}$ is the transformation matrix from moment space to velocity space, and its elements are determined by discrete velocity model(DVM). Once the discrete velocity model is determined, the form of matrix $\mathbf{C}$ is known. The choice of the discrete velocity depends on numerical efficiency, numerical stability, and to which extent the local symmetry should be kept. The last point relies on the specific physical problem under consideration. To capture the first order TNE behaviors, we adopt the D2V16 discrete velocity model. Sketches (a) and (b) of two kinds of D2V16 model are shown in Fig. \ref{fig1}. The specific values of sketch (a) and (b) are given in the following equations, respectively.

\[a:\mathbf{v}_{i}=(v_{ix},v_{iy})=
\left\{
\begin{array}{lll}
c[\textup{cos}\frac{(i-1)\pi}{2},\textup{sin}\frac{(i-1)\pi}{2}],     &i& = 1-4 \tt{,} \\
2c[\textup{cos}\frac{2(i-1)\pi}{4},\textup{sin}\frac{(2i-1)\pi}{4}],  &i& = 5-8 \tt{,} \\
3c[\textup{cos}\frac{(i-9)\pi}{2},\textup{sin}\frac{(i-9)\pi}{2}],    &i& = 9-12 \tt{,} \\
4c[\textup{cos}\frac{(2i-9)\pi}{4},\textup{sin}\frac{(2i-9)\pi}{4}],  &i& = 13-16 \tt{.}
\end{array} \label{Eq:DDBM-DVM1}
\right.
\]

\[b:\mathbf{v}_{i}=(v_{ix},v_{iy})=
\left\{
\begin{array}{lll}
cyc:c(\pm1,0),     &i& = 1-4 \tt{,} \\
cyc:c(\pm1,\pm1),  &i& = 5-8 \tt{,} \\
cyc:2c(\pm1,0),    &i& = 9-12 \tt{,} \\
cyc:2c(\pm1,\pm1),  &i& = 13-16 \tt{.}
\end{array}
\right.
\]
where $c$ is an adjustable parameter and ``cyc'' indicates the cyclic permutation. $\eta_{i}=\eta_{0}$ for $i=1-4$, and $\eta_{i}=0$ for $i=5-16$ in the two sketches of D2V16. The discrete form of $\mathbf{f}^{\sigma,ES}$ can be obtained as follow.
\begin{equation}
\mathbf{f}^{\sigma,ES}=\mathbf{C}^{-1}\mathbf{\hat{f}}^{\sigma,ES}
\tt{,}
\end{equation}
where $\mathbf{C}^{-1}$ is the inverse matrix of $\mathbf{C}$, which can be analytically solved by using some software, for example, MATLAB. The specific values of matrix $\mathbf{C}$ are referred to Ref.\cite{2016Lai-DBMRT}.

\subsection{Ellipsoidal Statistical BGK model and Navier-Stokes equations}\label{NS equation}

There are two kinds of ways to obtain the macroscopic fluid equations. The first is the traditional which is based on the continuum assumption and near equilibrium approximation. The second is to start from kinetic theory and obtain through some multiscale analysis method, such as the Chapman-Enskog expansion.  For the convenience of description, the second way to obtain macroscopic fluid equations is referred to Kinetic Macroscopic Modeling method. In contrast, the discrete Boltzmann modeling method is a Kinetic Direct Modeling method.

In this work, the proposed DBM has two physics functions. The first is to recover hydrodynamic NS equations in the continuum limit, which can be proved by using the Chapman-Enskog analysis. The second is to present various thermodynamic non-equilibrium behaviors. Based on one-step model, we can present NS equations with different forms by performing CE expansion with different local equilibrium distribution function.
The discrete Boltzmann equation can be written as
\begin{equation}
\frac{\partial f^{\sigma}_{i}}{\partial t}+v_{i\alpha}\cdot\frac{\partial f^{\sigma}_{i}}{\partial r_{i\alpha}}=-\frac{1}{\tau^{\sigma}}(f^{\sigma}_{i}-f^{\sigma,ES}_{i})\tt{,}
\label{Eq:DDBM-N19}
\end{equation}
where $f^{\sigma,ES}=f^{\sigma,ES}(\rho^{\sigma},\mathbf{u},T)$. In this model $f^{\sigma}$ tends to $f^{\sigma,ES}$ directly. The CE expansion is performed around the equilibrium distribution function of component $\sigma$ ,
\begin{equation}
f^{\sigma,meq}_{i}=f^{\sigma,meq}_{i} (\rho^{\sigma}, \mathbf{u}, T)
\tt{,}      \label{Eq:DDBM-N20a}
\end{equation}
which depends on the density of component $\sigma$, flow velocity and temperature of the mixture. The velocity distribution functions can be expanded as:
\begin{equation}
f^{\sigma}_{i}=f^{\sigma,m eq}_{i}+\epsilon f^{\sigma,(1)}_{i}+\epsilon^{2}f^{\sigma,(2)}_{i}+\cdots
\tt{,}      \label{Eq:DDBM-N20}
\end{equation}
\begin{equation}
f^{\sigma,ES}_{i}=f^{\sigma,m eq}_{i}+\epsilon f^{\sigma,ES(1)}_{i}+\epsilon^{2}f^{\sigma,ES(2)}_{i}+\cdots
\tt{,}   \label{Eq:DDBM-N21}
\end{equation}
where $\epsilon$ is a coefficient referring to Knudsen number, the partial derivative of time and space can also be expanded to
\begin{equation}
\frac{\partial}{\partial t}=\epsilon\frac{\partial}{\partial t_{1}}+\epsilon^{2}\frac{\partial}{\partial t}_{2}+\cdots
\tt{,}      \label{Eq:DDBM-N22}
\end{equation}
\begin{equation}
\frac{\partial}{\partial r_{\alpha}}=\epsilon\frac{\partial}{\partial r_{1\alpha}}
\tt{,}       \label{Eq:DDBM-N23}
\end{equation}
By substituting the Eqs. (\ref{Eq:DDBM-N20}) - (\ref{Eq:DDBM-N23}) into Eq. (\ref{Eq:DDBM-N19}), we obtain the NS equations as follows:
\begin{equation}
\frac{\partial\rho^{\sigma}}{\partial t}+\frac{\partial}{\partial r_{\alpha}}(\rho^{\sigma} u_{\alpha})=0
\tt{,}
\end{equation}
\begin{eqnarray}
\begin{aligned}
\frac{\partial}{\partial t}(\rho^{\sigma} u_{\alpha})
&+\frac{\partial (p^{\sigma}\delta_{\alpha\beta}+\rho^{\sigma}
u_{\alpha}u_{\beta})}{\partial r_{\beta}}+\frac{\partial P^{\sigma}_{\alpha\beta}}{\partial r_{\beta}}=0
\end{aligned}
\end{eqnarray}
\begin{eqnarray}
\begin{aligned}
\frac{\partial}{\partial t}\rho^{\sigma}E^{\sigma}_{T}&+\frac{\partial}{\partial r_{\alpha}}(\rho^{\sigma}E^{\sigma}_{T}+p^{\sigma})u_{\alpha}
+\frac{\partial}{\partial r_{\beta}}[u_{\alpha}P^{\sigma}_{\alpha\beta}\\
&-\kappa^{\sigma}\frac{\partial (T/m^{\sigma})}{\partial r_{\alpha}}]=0
\tt{.}
\end{aligned}
\end{eqnarray}
\begin{equation}
P^{\sigma}_{\alpha\beta}=-\mu^{\sigma}(\frac{\partial u_{\alpha}}{\partial r_{\beta}}+\frac{\partial u_{\beta}}{\partial r_{\alpha}}-\frac{2}{D}\frac{\partial u_{\gamma}}{\partial r_{\gamma}}\delta_{\alpha\beta})
\tt{,}
\end{equation}
where $p^{\sigma}=n^{\sigma}kT$, $E^{\sigma}_{T}=\frac{1}{2}[D*T/m^{\sigma}+u^{2}_{\alpha}]$, $\mu^{\sigma}=\frac{1}{1-b}\tau^{\sigma} p^{\sigma}$, $\kappa^{\sigma}=C_{p}^{\sigma}\tau^{\sigma} p^{\sigma}$ are the pressure, the energy per unit mass, the dynamic viscosity coefficient, and heat conductivity of species $\sigma$, respectively. $k$ represents the Boltzmann constant. $C_{p}^{\sigma}$ represents specific heat at constant pressure and $C_{p}^{\sigma}=\frac{D+2}{2}R$. The Prandtl number, $\Pr^{\sigma}=\frac{C_{p}\mu^{\sigma}}{\kappa^{\sigma}}=\frac{1}{1-b}$, is flexible with the coefficient $b$. For convenience, we do not consider extra degree of freedom in the above and following derivations.

As we can see, the NS equations above do not show explicitly the inter-component interaction, including the inter-component diffusion, inter-component heat conduction, etc. In order to include explicitly the inter-component interaction in NS equations, we use a second way in recovering NS equations.
The discrete Boltzmann equation is re-written as
\begin{equation}
\frac{\partial f^{\sigma}_{i}}{\partial t}+v_{i\alpha}\cdot\frac{\partial f^{\sigma}_{i}}{\partial r_{i\alpha}}=-\frac{1}{\tau^{\sigma}_{1}}(f^{\sigma}_{i}-f^{\sigma,sES}_{i})-\frac{1}{\tau^{\sigma}_{2}}(f^{\sigma,sES}_{i}-f^{\sigma,ES}_{i})\tt{.}
\end{equation}
with $\tau^{\sigma}_{1}=\tau^{\sigma}_{2}=\tau^{\sigma}$. For convenience of description, we define
\[ S_{i}^{\sigma}=\frac{1}{\tau^{\sigma}}(f^{\sigma,sES}-f^{\sigma,ES})\]
where $f^{\sigma,sES}=f^{\sigma,sES}(\rho^{\sigma},\mathbf{u}^{\sigma},T^{\sigma})$ and $f^{\sigma,ES}=f^{\sigma,ES}(\rho^{\sigma},\mathbf{u},T)$. The CE expansion is performed around the equilibrium distribution function of component $\sigma$,
\begin{equation}\label{}
f^{\sigma,eq}_{i}=f^{\sigma,eq}_{i}(\rho^{\sigma}, \mathbf{u}^{\sigma}, T^{\sigma})
\tt{,}
\end{equation}
which depends on the density, flow velocity and temperature of component $\sigma$.
The velocity distribution function can be expanded as:
\begin{equation}
f^{\sigma}_{i}=f^{\sigma,eq}_{i}+\epsilon f^{\sigma,(1)}_{i}+\epsilon^{2}f^{\sigma,(2)}_{i}+\cdots
\tt{,}
\end{equation}
\begin{equation}
f^{\sigma,sES}_{i}=f^{\sigma,eq}_{i}+\epsilon f^{\sigma,sES(1)}_{i}+\epsilon^{2}f^{\sigma,sES(2)}_{i}+\cdots
\tt{,}
\end{equation}
\begin{equation}
f^{\sigma,ES}_{i}=f^{\sigma,meq}_{i}+\epsilon f^{\sigma,ES(1)}_{i}+\epsilon^{2}f^{\sigma,ES(2)}_{i}+\cdots
\tt{,}
\end{equation}
and $S_{i}^{\sigma}=\epsilon S_{i}^{\sigma}$.
Via CE analysis, this model can recover to the NS equations in the hydrodynamic limit as follows:
\begin{equation}
\frac{\partial\rho^{\sigma}}{\partial t}+\frac{\partial}{\partial r_{\alpha}}(\rho^{\sigma} u^{\sigma}_{\alpha})=0
\tt{,} \label{Eq:DDBM-N2}
\end{equation}
\begin{eqnarray}
\begin{aligned}
\frac{\partial}{\partial t}(\rho^{\sigma} u^{\sigma}_{\alpha})
&+\frac{\partial (p^{\sigma}\delta_{\alpha\beta}+\rho^{\sigma}
u^{\sigma}_{\alpha}u^{\sigma}_{\beta})}{\partial r_{\beta}}+\frac{\partial (P^{\sigma}_{\alpha\beta}+U_{\alpha\beta}^{\sigma})}{\partial r_{\beta}}\\
&=-\frac{\rho^{\sigma}}
{\tau^{\sigma}}(u_{\alpha}^{\sigma}-u_{\alpha})\tt{,}\label{Eq:DDBM-N3}
\end{aligned}
\end{eqnarray}
\begin{eqnarray}
\begin{aligned}
\frac{\partial}{\partial t}\rho^{\sigma}E^{\sigma}_{T}&+\frac{\partial}{\partial r_{\alpha}}(\rho^{\sigma}E^{\sigma}_{T}+p^{\sigma})u^{\sigma}_{\alpha}
+\frac{\partial}{\partial r_{\beta}}[u^{\sigma}_{\alpha}(P^{\sigma}_{\alpha\beta}+U^{\sigma}_{\alpha\beta})\\
&-\kappa^{\sigma}\frac{\partial (T^{\sigma}/m^{\sigma})}{\partial r_{\alpha}}+Y_{\beta}^{\sigma}]
=-\frac{\rho^{\sigma}}{\tau^{\sigma}}[D*\frac{T^{\sigma}-T}{2m^{\sigma}}\\
&+\frac{1}{2}(u^{\sigma2}_{\alpha}-u^{2}_{\alpha})]
\tt{.}\label{Eq:DDBM-N4}
\end{aligned}
\end{eqnarray}
with
\begin{equation}
P^{\sigma}_{\alpha\beta}=-\mu^{\sigma}(\frac{\partial u^{\sigma}_{\alpha}}{\partial r_{\beta}}+\frac{\partial u^{\sigma}_{\beta}}{\partial r_{\alpha}}-\frac{2}{D}\frac{\partial u^{\sigma}_{\gamma}}{\partial r_{\gamma}}\delta_{\alpha\beta})
\tt{,}
\end{equation}
\begin{equation}
U_{\alpha\beta}^{\sigma}=\frac{1}{1-b}\rho^{\sigma}[(u_{\beta}-u^{\sigma}_{\beta})(u_{\alpha}-u^{\sigma}_{\alpha})
+\frac{1}{D}(u^{\sigma}_{\alpha}-u_{\alpha})^{2}\delta_{\alpha\beta}]
\end{equation}
\begin{eqnarray}
\begin{aligned}
Y_{\alpha}^{\sigma}&=[\frac{D}{2m^{\sigma}}\rho^{\sigma}k(T^{\sigma}-T)(u^{\sigma}_{\alpha}-u_{\alpha})\\
&-\frac{1}{D}\rho^{\sigma}(u_{\alpha}^{\sigma}-u_{\alpha})^2u_{\alpha}^{\sigma}\\
&+\rho^{\alpha}u_{\alpha}^{\sigma}(u_{\alpha}^{\sigma}-u_{\alpha})u_{\alpha}^{\sigma}]+\frac{1}{2}\rho^{\sigma}(u^{\sigma2}_{\alpha}-u_{\alpha}^{2})(u_{\alpha}^{\sigma}-u_{\alpha})
\end{aligned}
\end{eqnarray}
where $p^{\sigma}=n^{\sigma}kT^{\sigma}$, $E^{\sigma}_{T}=\frac{1}{2}[D*T^{\sigma}/m^{\sigma}+u^{\sigma 2}]$, $\mu^{\sigma}=\frac{1}{1-b}\tau^{\sigma} p^{\sigma}$, $\kappa^{\sigma}=C_{p}^{\sigma}\tau^{\sigma} p^{\sigma}$ are the pressure, the energy per unit mass, the dynamic viscosity coefficient, and heat conductivity of species $\sigma$, respectively. $C_{p}^{\sigma}=\frac{D+2}{2}R$ and $\Pr^{\sigma}=\frac{C_{p}^{\sigma}\mu^{\sigma}}{\kappa^{\sigma}}=\frac{1}{1-b}$.
The right items of the equal sign of equations (\ref{Eq:DDBM-N3}) and (\ref{Eq:DDBM-N4}) represent the momentum exchange and energy exchange between two components, which is sourced from particles collision. Although different in forms between two set of NS equations, they are all right physical. The former is more coarse-grained.

Performing the operator $\sum\limits_{\sigma}$ to the two sides of Eqs. (\ref{Eq:DDBM-N2})-(\ref{Eq:DDBM-N4}) gives the NS equations describing the whole system.
\begin{equation}
\frac{\partial\rho}{\partial t}+\frac{\partial}{\partial r_{\alpha}}(\rho u_{\alpha})=0
\tt{,}
\end{equation}
\begin{eqnarray}
\begin{aligned}
\frac{\partial}{\partial t}(\rho u_{\alpha})
&+\frac{\partial \sum\limits_{\sigma}(p^{\sigma}\delta_{\alpha\beta}+\rho^{\sigma}
u^{\sigma}_{\alpha}u^{\sigma}_{\beta})}{\partial r_{\beta}}\\
&+\frac{\partial\sum\limits_{\sigma} (P^{\sigma}_{\alpha\beta}+U_{\alpha \beta}^{\sigma})}{\partial r_{\beta}}=0\tt{,} \label{Eq:DDBM-N1}
\end{aligned}
\end{eqnarray}
\begin{eqnarray}
\begin{aligned}
\frac{\partial}{\partial t}\rho E_{T}&+\frac{\partial}{\partial r_{\alpha}}\sum\limits_{\sigma}(\rho^{\sigma}E^{\sigma}_{T}+p^{\sigma})u^{\sigma}_{\alpha}\\
&-\frac{\partial}{\partial r_{\beta}}\sum\limits_{\sigma}[u^{\sigma}_{\beta}(P^{\sigma}_{\alpha\beta}+U_{\alpha \beta}^{\sigma})-\kappa^{\sigma}\frac{\partial (T^{\sigma}/m^{\sigma})}{\partial r_{\alpha}}+Y_{\alpha}^{\sigma}]=0 \tt{.}
\end{aligned}
\end{eqnarray}
When the temperature and velocity of each component approach the same, we have $T^{\sigma}=T$ and $u^{\sigma}= u$. The Eq. (\ref{Eq:DDBM-N1}) is equivalent to
\begin{eqnarray}
\begin{aligned}
\frac{\partial}{\partial t}(\rho u_{\alpha})+\frac{\partial}{\partial r_{\beta}}(p\delta_{\alpha \beta}+\rho u_{\alpha}u_{\beta})+\frac{\partial P_{\alpha\beta}}{\partial r_{\beta}}=0
\end{aligned}
\end{eqnarray}
where
\begin{equation}
p=\sum\limits_{\sigma}p^{\sigma}=\sum\limits_{\sigma}n^{\sigma}T^{\sigma}
\tt{,}
\end{equation}
\begin{equation}
P_{\alpha \beta}=\sum\limits_{\sigma}P_{\alpha \beta}^{\sigma}=-\mu(\frac{\partial u_{\alpha}}{\partial r_{\beta}}+\frac{\partial u_{\beta}}{\partial r_{\alpha}}-\frac{2}{D}\frac{\partial u_{\gamma}}{\partial r_{\gamma}}\delta_{\alpha\beta})
\tt{,}
\end{equation}
with the dynamic viscosity coefficient of the whole system
\begin{equation}
\mu=\sum\limits_{\sigma}\mu^{\sigma}=\frac{1}{1-b}\sum\limits_{\sigma}(p^{\sigma}\tau^{\sigma})=\frac{1}{1-b}p\tau
\end{equation}
The heat conductivity of the whole system is
\begin{equation}
\kappa=\sum\limits_{\sigma}\kappa^{\sigma}=\sum\limits_{\sigma}(C_{p}^{\sigma}\tau^{\sigma}p^{\sigma})
\end{equation}
The specific heat at constant pressure of the whole system is
\begin{equation}
C_{p}=\frac{\sum\limits_{\sigma}n^{\sigma} C_{p}^{\sigma}}{\sum\limits_{\sigma}n^{\sigma}}
\end{equation}
 In addition, it is easy to demonstrate the diffusion equations from NS equation\cite{2016Lin-CNF,Xu2005PRE}. Shown in the following equations are Fick's first law, Fick's second law, and Stefan-Maxwell diffusion equation, respectively.
\begin{equation}
J^{\sigma}_{\alpha}=-D^{\sigma}_{d}\frac{\partial \rho^{\sigma}}{\partial r_{\alpha}}
\end{equation}
\begin{equation}
\frac{\partial \lambda^{\sigma}}{\partial t}=D^{\sigma}_{d}\frac{\partial}{\partial r}(\frac{\partial \lambda^{\sigma}}{\partial r})
\end{equation}
\begin{equation}
M^{A}M^{B}(u^{B}_{\alpha}-u^{A}_{\alpha})=D_{d}\frac{\partial M^{A}}{\partial r_{\alpha}}-D_{d}(\lambda ^{A}-M^{A})\frac{1}{p}\frac{p}{\partial r_{\alpha}}
\end{equation}
where $J^{\sigma}_{\alpha}=\rho^{\sigma}(u^{\sigma}_{\alpha}-u_{\alpha})$ is the is the diffusive flux of mass, $D_{d}^{\sigma}=\tau^{\sigma}T^{\sigma}/m^{\sigma}$ is the diffusivity of components $\sigma$, $\lambda^{\sigma}$ is the mass fraction of $\sigma$, $M^{\sigma}$ the mole fraction, and
\begin{equation}
D_{d}=\frac{\rho}{\rho^{A}\rho^{B}}M^{A}M^{B}p\tau
\end{equation}
the diffusion coefficient of the whole system. The heat transform of component A is
\begin{equation}
j_{q}^{A}=\frac{D}{2}n^{A}n^{B}\frac{kT^{A}-kT^{B}}{n^{A}+n^{B}}
\end{equation}
It should be noted that the role of CE analysis in DBM modeling is only to facilitate query and validate the kinetic moment relations that need to be preserved. Whether or not to finish the derivation to obtain the final hydrodynamic equations does not affect DBM modeling and simulation. Such a modeling method is valid under the condition that the Knudsen number is not too large so that the CE expansion theory still works. A second point to be noted is that the DBM obtained via the kinetic direction modeling method is unique, while the macroscopic models, described by fluid equations, obtained via the kinetic macroscopic modeling method may be different. That is to say, the correspondence between the macroscopic fluid model and the DBM may be several-to-one.

\section{Numerical simulations}\label{Numerical simulations}

In this section, five types of validations and verifications of the two-fluid DBM with a flexible Prandtl number are performed. The first validation is a one-dimensional binary diffusion problem in isothermal condition. The second is 1-dimensional Riemann problems for compressible flows with high Mach number. The third is a two-dimensional KH instability simulation. The fourth is a two-dimensional regular reflection of a shock wave and the fifth the two-dimensional shock wave act on a cylindrical bubble. The first sketch of D2V16 model is adopted in all simulations except that the fourth simulation where the second sketch is used, because of the better numerical stability of the second sketch in the fourth simulation. In addition, the first order forward difference scheme and the second order nonoscillatory nonfree dissipative(NND) scheme are used to discrete the temporal and spatial derivatives \cite{2016Lai-DBMRT,Zhang1992NND}, respectively. Besides the validations and verifications, some interesting TNE behaviors, which are not available a NS model or a single-fluid DBM, are presented.

\subsection{Binary diffusion}

\begin{figure}
\begin{center}
\includegraphics[width=0.4\textwidth]{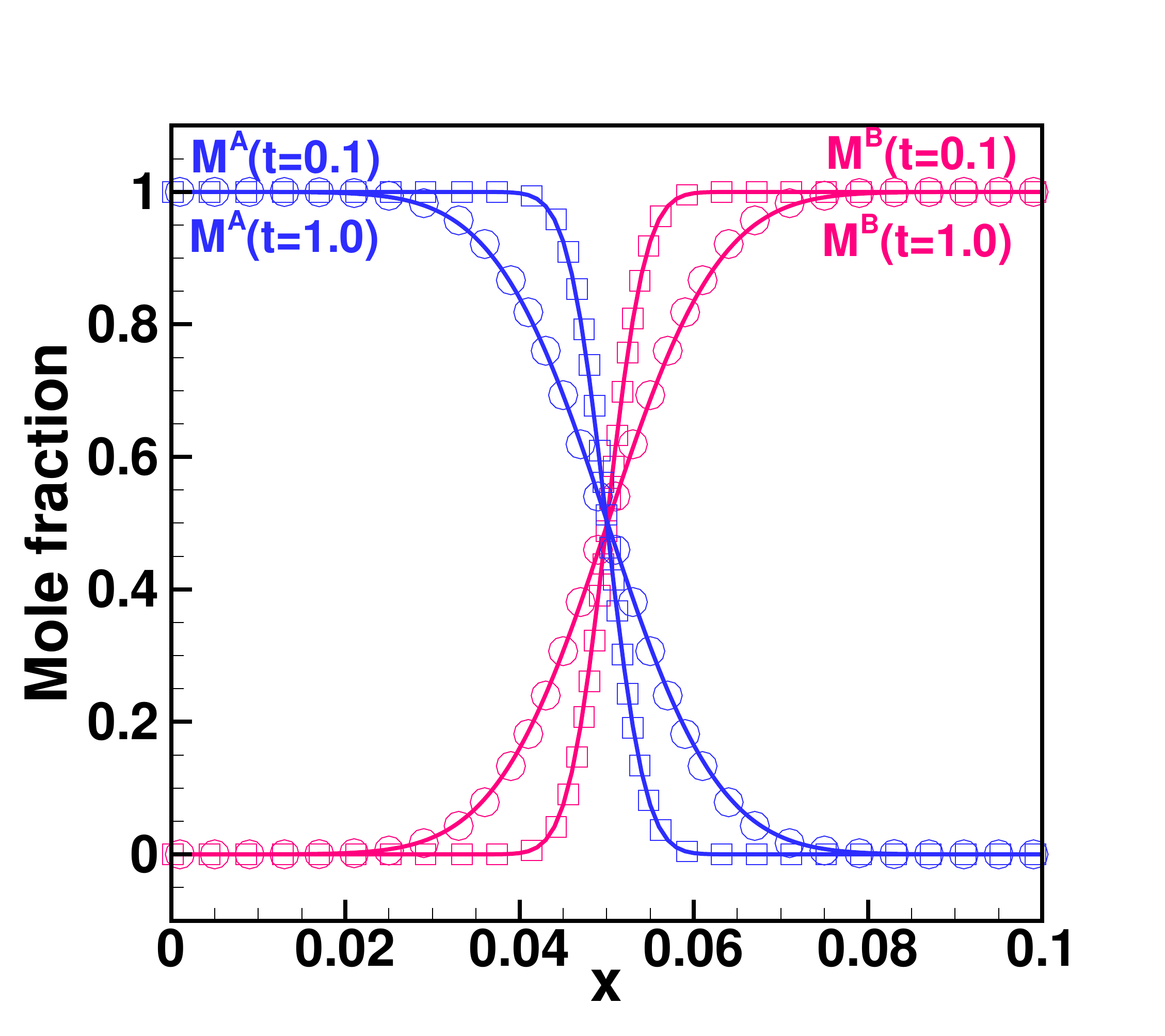}
\end{center}
\caption{ Mole fractions $M^{A}$(blue line) and $M^{B}$(red line) in the binary diffusion at two instants: $t=0.1$ and $1.0$, respectively. Symbols denote DBM simulation results and continuous lines denote the corresponding analytical solutions.} \label{fig2}
\end{figure}

Diffusions take place in a system when two miscible species contact each other. Diffusion is a common phenomenon in the nature and engineering, its evolution of macroscopic concentration for each species can be described by the Fick's law in isothermal condition\cite{2016Lin-CNF}. The following analytical solution works
\begin{equation}
M^{\sigma}=\frac{1}{2}+\frac{\Delta M^{\sigma}}{2} \tt{erf}(\frac{x}{\sqrt{4D_{d}t}})
\tt{,}
\end{equation}
where $\Delta M^{\sigma}$ is the initial mole fraction difference and $D_{d}$ the diffusion coefficient. For comparing with this solution, we simulate an isothermal diffusion here. The mixture of two gases is initially given by the following step function:
\[
\left\{
\begin{array}{l}
(M^{A},M^{B})_{L}=(100\%,0\%) \tt{,} \\
(M^{A},M^{B})_{R}=(0\%,100\%) \tt{,}
\end{array}
\right.
\]
where the suffix $L$ indexes the left part and $R$ the right part along the horizontal direction $x$. The molecular masses $m_{A}=1$, $m_{B}=1$, the relaxation time $\tau^{A}=\tau^{B}=5\times10^{-5}$, and the other parameters $c=1.0$, $I^{A}=I^{B}=3$, $\Delta t =1\times10^{-5}$, $\Delta x=\Delta y=2\times10^{-4}$, $\eta^{A}=\eta^{B}=10$, $b=0$, $N_{x}\times N_{y}=500\times1$. The zero gradient boundary condition is adopted in the $x$ direction.
Figure \ref{fig2} shows the comparison of mole fraction of two components between the DBM simulation results and the analytical solutions, with $\Delta M^{\sigma}=1.0$ and $D_{d}=0.001$. The analytical solutions are denoted by solid lines, and the corresponding simulation results at constants $t=0.1$ and $t=1.0$ are denoted by squares and circles, respectively. Figure 2 shows the satisfying agreements between results of simulation and analysis. It is confirmed that DBM can precisely describe the interaction of two components.
Besides, it can be noticed that the parameters of the two components can be set to be equal or not, which can not achieve on a single-fluid DBM.

\subsection{Riemann problems}

It is well known that the Riemann problems are classical problems to verify the ability of a model to capture shock wave. Simply, the Riemann problems can be seen as single-fluid problem. However, in this section, our 2-dimensional 2-fluid DBM is used to solve the 1-dimensional Riemann problems. Besides giving the results that can be obtained from single-fluid DBM, we can get more accurate physical information than single-fluid DBM. Now, we give simulation results for four typical Riemann problems, i.e., the Sod's shock tube, the Lax's shock tube, the Sjogreen's problem, and the collision of two strong shock waves. In addition, we simulate a Sod' shock tube of two components with different particle masses, which can not be achieved from the simple Riemann analytical solution or a single-fluid DBM. Initially, the flow field with grid $N_{x}\times N_{y}=1000\times1$ is equally divided into left side ``A'' and right side ``B'', ``A'' and ``B'' are indexes of fluid components. And there is just component A on the ``A'' side and component B on the ``B'' side for all five problems, respectively. The major initial conditions of the flow field are shown by Table (\ref{table1}). And, we adopt the zero gradient boundary condition in the $x$ direction for all the five problems.
\begin{center}
\begin{table*}[htbp]
\begin{tabular}{|p{3cm}<{\centering} |c| p{7cm}|}
\hline
Items& Pr number and particle mass &Initial condition\\
\hline
1.Sod's shock tube& \multirow{2}{*}\centering{Pr=0.8,1.0,2.0}{($m^{A}=m^{B}=1$)}&
$\left\{
\begin{array}{l}
(\rho,T,U_{x},U_{y})_{A}=(1.0,1.0,0.0,0.0)\\
(\rho,T,U_{x},U_{y})_{B}=(0.125,0.8,0.0,0.0)
\end{array}
\right.$\\
\hline
2.Lax's shock tube& \multirow{2}{*}\centering{Pr=0.8,1.0,2.0}{($m^{A}=m^{B}=1$)}&
$\left\{
\begin{array}{l}
(\rho,T,U_{x},U_{y})_{A}=(0.445,7.928,0.698,0.0) \\
(\rho,T,U_{x},U_{y})_{B}=(0.5,1.142,0.0,0.0)
\end{array}
\right.$\\
\hline
3.Sjogreen's problem& \multirow{2}{*}\centering{Pr=1.0}{($m^{A}=m^{B}=1$)}&
$\left\{
\begin{array}{l}
(\rho,T,U_{x},U_{y})_{A}=(1.0,0.5,-1.2,0.0) \\
(\rho,T,U_{x},U_{y})_{B}=(1.0,0.5,1.2,0.0)
\end{array}
\right.$\\
\hline
4.The collision of two strong shock waves& \multirow{2}{*}\centering{Pr=0.8,1.0,2.0}{($m^{A}=m^{B}=1$)}&
$\left\{
\begin{array}{l}
(\rho,T,U_{x},U_{y})_{A}=(5.99924,76.8254,19.5975,0.0) \\
(\rho,T,U_{x},U_{y})_{B}=(5.99242,7.69222,-6.19633,0.0)
\end{array}
\right.$\\
\hline
5.Sod's shock tube with different particle masses& \multirow{2}{*}\centering{Pr=1.0}{($m^{B}=3m^{A}=3$)}&
Expect for particle mass, the initial conditions are the same with Sod's shock tube.\\
\hline
\hline
\end{tabular}
\caption{The main initial conditions of flow field of Riemann problems, respectively.}
\label{table1}
\end{table*}
\end{center}

\subsubsection{Sod's shock tube}\label{Sod' shock tube}

\begin{figure}
\begin{center}
\includegraphics[width=0.5\textwidth]{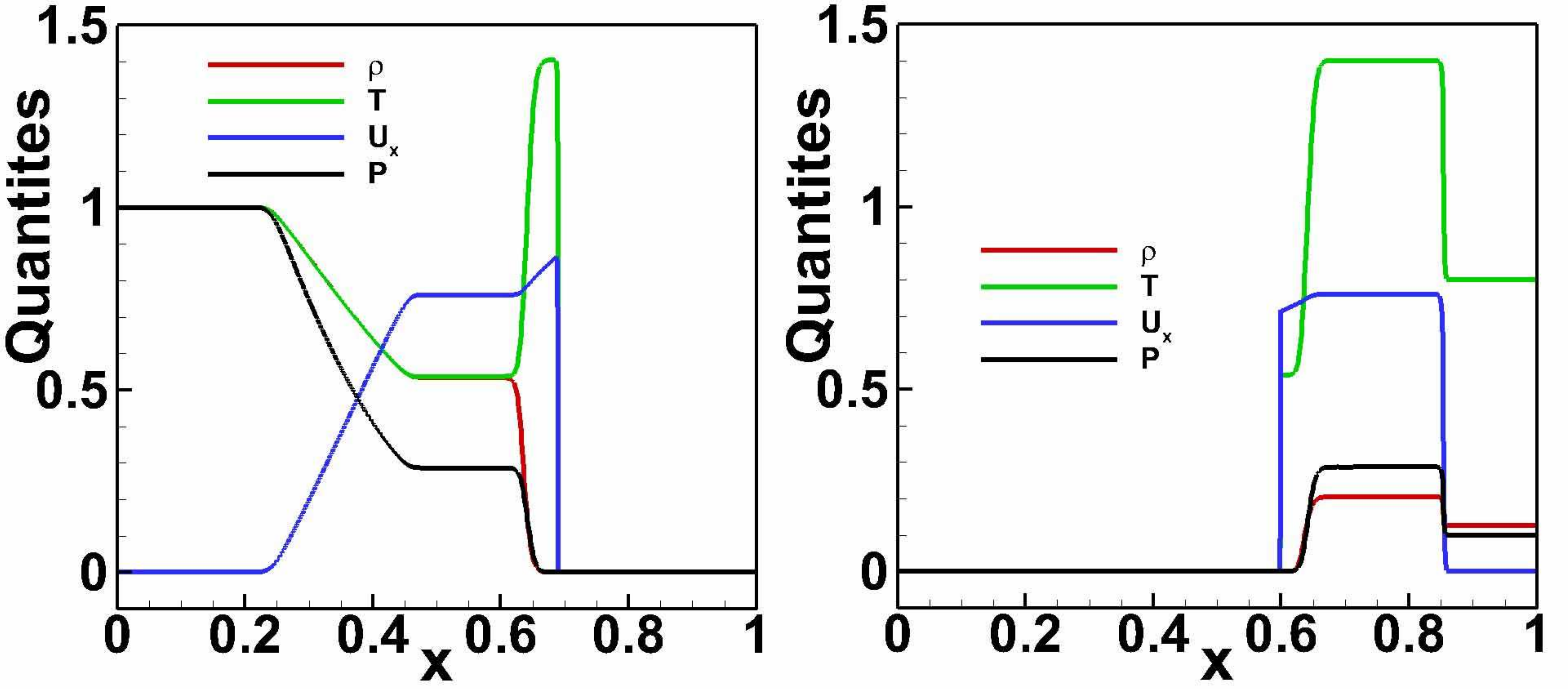}
\end{center}
\caption{Profiles of density (red line), temperature (green line), $U_{x}$ (blue line) and pressure (black line) of component A(left picture) and B(right picture) at $t=0.18$, with $\Pr=1.0$, respectively. Such a result is not available from a single-fluid model.} \label{fig3}
\end{figure}

\begin{figure}
\begin{center}
\includegraphics[width=0.4\textwidth]{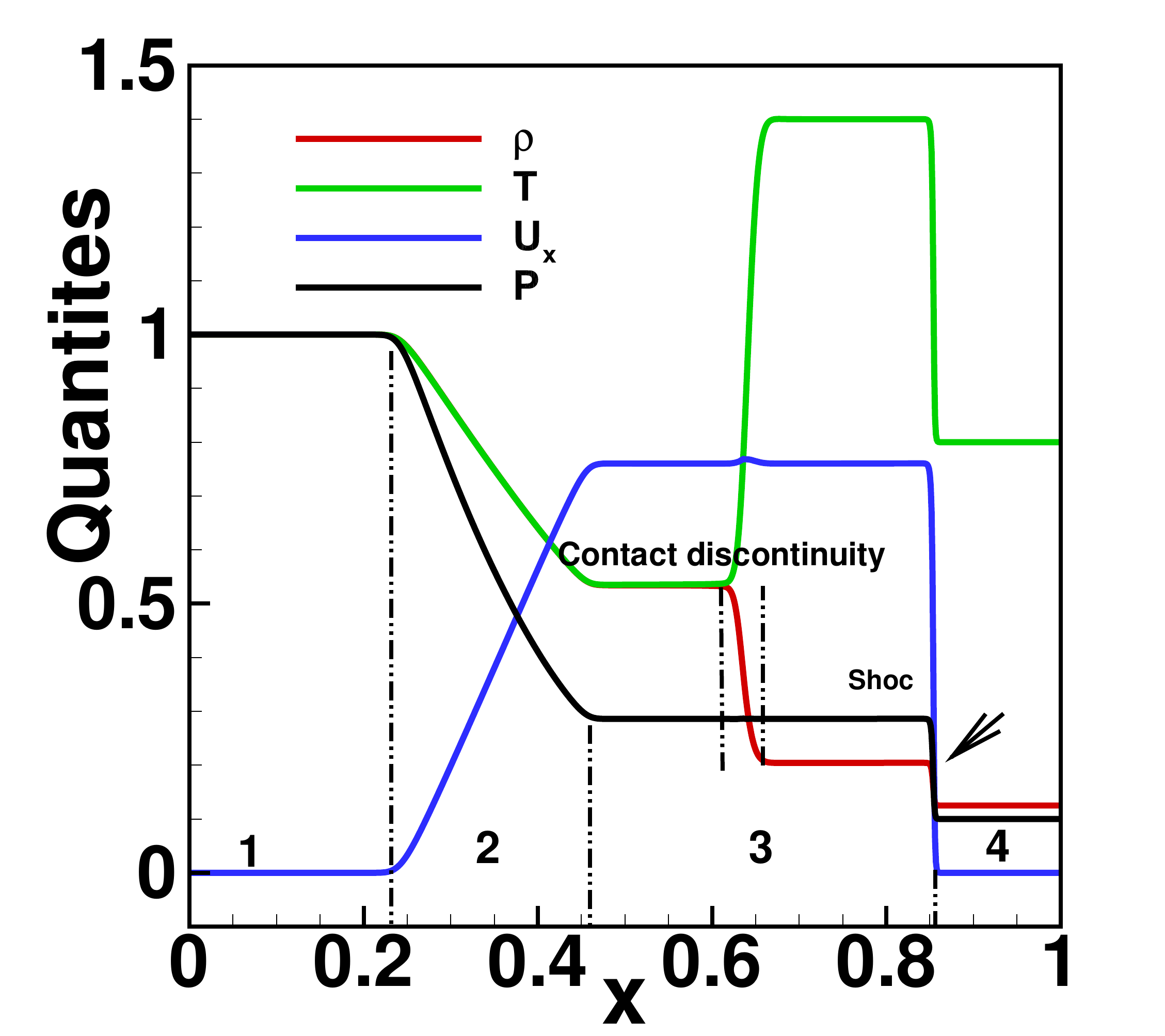}
\end{center}
\caption{Profiles of density (red line), temperature (green line), $U_{x}$ (blue line) and pressure (black line) of the physical system at $t=0.18$, with $\Pr=1.0$, respectively.} \label{fig4}
\end{figure}

The initial conditions of other quantities are $c=1.0$, $m^{A}=m^{B}=1$, $\tau^{A}=\tau^{B}=2\times 10^{-4}$, $I^{A}=I^{B}=0(\gamma^{A}=\gamma^{B}=2.0)$, $b=0.0(\Pr=1.0)$, $\Delta t=2\times10^{-6}$, $\Delta x=\Delta y=10^{-3}$, $\eta^{A}=\eta^{B}=0$.
As we mentioned above, a two-fluid DBM can provides more accurate physical information than a single-fluid DBM. For example, as shown in Fig. \ref{fig3}, a two-fluid DBM can provide the profiles of quantities of each component, which can not be obtained from a single-fluid DBM.
Besides, we can also obtain the quantities profiles of the physical system. The profiles of density, temperature, velocity, and pressure of the physical system at $t=0.18$  with $\Pr=1.0$ are shown in the Fig. \ref{fig4}. Meanwhile, it is clear that the Sod's tube can be divided into four parts: Part 1 and Part 4 the undisturbed area, Part 2 the left-propagating rarefaction wave and Part 3 the disturbed area.

\subsubsubsection{Comparison with analytical solution}

Here we quantitatively compare the DBM results with the analytical solution. As shown by Fig. \ref{fig4}, a shock wave can be seen easily between Part 3 (wave rear) and Part 4 (wavefront) and the quantities of the two sides are satisfied with Rankine-Hugoniot conditions as follows:

\begin{equation}
\frac{u}{c_{0}}=\frac{2}{\gamma +1}(\tt{Ma}-\frac{1}{\tt{Ma}})+\frac{u_{0}}{c_{0}} \label{Eq:DDBM-RH1}
\tt{,}
\end{equation}
\begin{equation}
\frac{p}{p_{0}}=\frac{2 \gamma}{\gamma +1}\tt{Ma}^{2}-\frac{\gamma -1}{\gamma +1} \label{Eq:DDBM-RH2}
\tt{,}
\end{equation}
\begin{equation}
\frac{\rho}{\rho_{0}}=\frac{(\gamma +1)\tt{Ma}^{2}}{(\gamma -1)\tt{Ma}^{2}+2} \label{Eq:DDBM-RH3}
\tt{,}
\end{equation}
where subscript ``0'' represents the wavefront (undisturbed area), and $c_{0}=\sqrt{\gamma T_{0}}$ the sound speed, $\tt{Ma}$ the mach number. The quantities of wavefront and wave rear are $(\rho_{0},p_{0},u_{0})=(0.125,0.1,0.0)$ and $u=0.76043$, respectively. By submitting $u_{0}$, $c_{0}$, $u$, and $\gamma$ into Eq. (\ref{Eq:DDBM-RH1}), we obtain $\tt{Ma}$=1.54783.
The simulation results are $(\rho,p)_{simulation}=(0.20435,0.28610)$, which are consistent with analytical solutions $(\rho,p)_{analysis}=(0.20438,0.28610)$ that calculated by Eqs. (\ref{Eq:DDBM-RH2}) and (\ref{Eq:DDBM-RH3}), indicating the ability of capturing 1-dimensional shock front accurately. In addition, a contact discontinuity can be seen in Part 3, which is continuous at profiles of pressure and velocity but discontinuous at density and temperature.

\begin{figure}
 \begin{center}
 \includegraphics[width=0.5\textwidth]{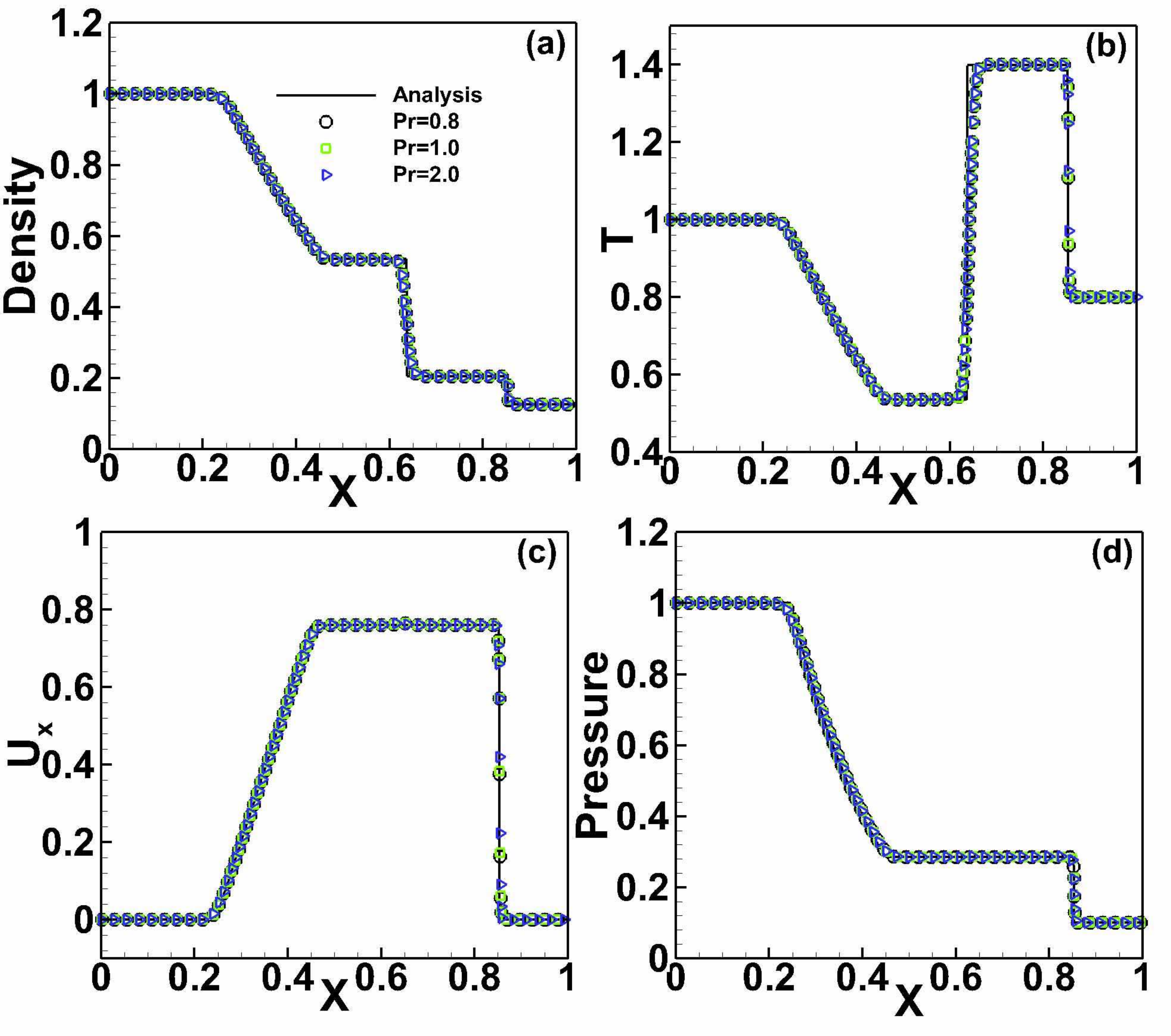}
 \end{center}
 \caption{Profiles of density (a), temperature (b), $U_{x}$ (c) and pressure (d) of the Sod' shock tube, at $t=0.18$. The lines indicate Riemann solutions, and the simulation results are
 denoted by circles, squares, and triangles, corresponding to Prandtl number $\Pr$ = 0.8, 1.0, and 2.0, respectively.} \label{fig5}
 \end{figure}

Shown in Fig. \ref{fig5} are the comparisons between simulation results and Riemann solutions of density, temperature, velocity, and pressure profiles at $t=0.18$, with the coefficient $b = -0.25$, $0.0$, and $0.5$, respectively (corresponding to Prandtl number $\Pr$ = 0.8, 1.0, and 2.0, respectively). It is clear that the left-propagating rarefaction wave and the right-propagating shock wave are captured by DBM. There exit distinct transition zones around the contact discontinuities for all four density, temperature, velocity, and pressure profiles. Because the Riemann solutions are based on Euler equations, which does not include the effects of viscosity and heat flux, but the DBM contains those. So the DBM results have smooth transition zones while Riemann solutions do not.

\begin{figure}
\begin{center}
\includegraphics[width=0.4\textwidth]{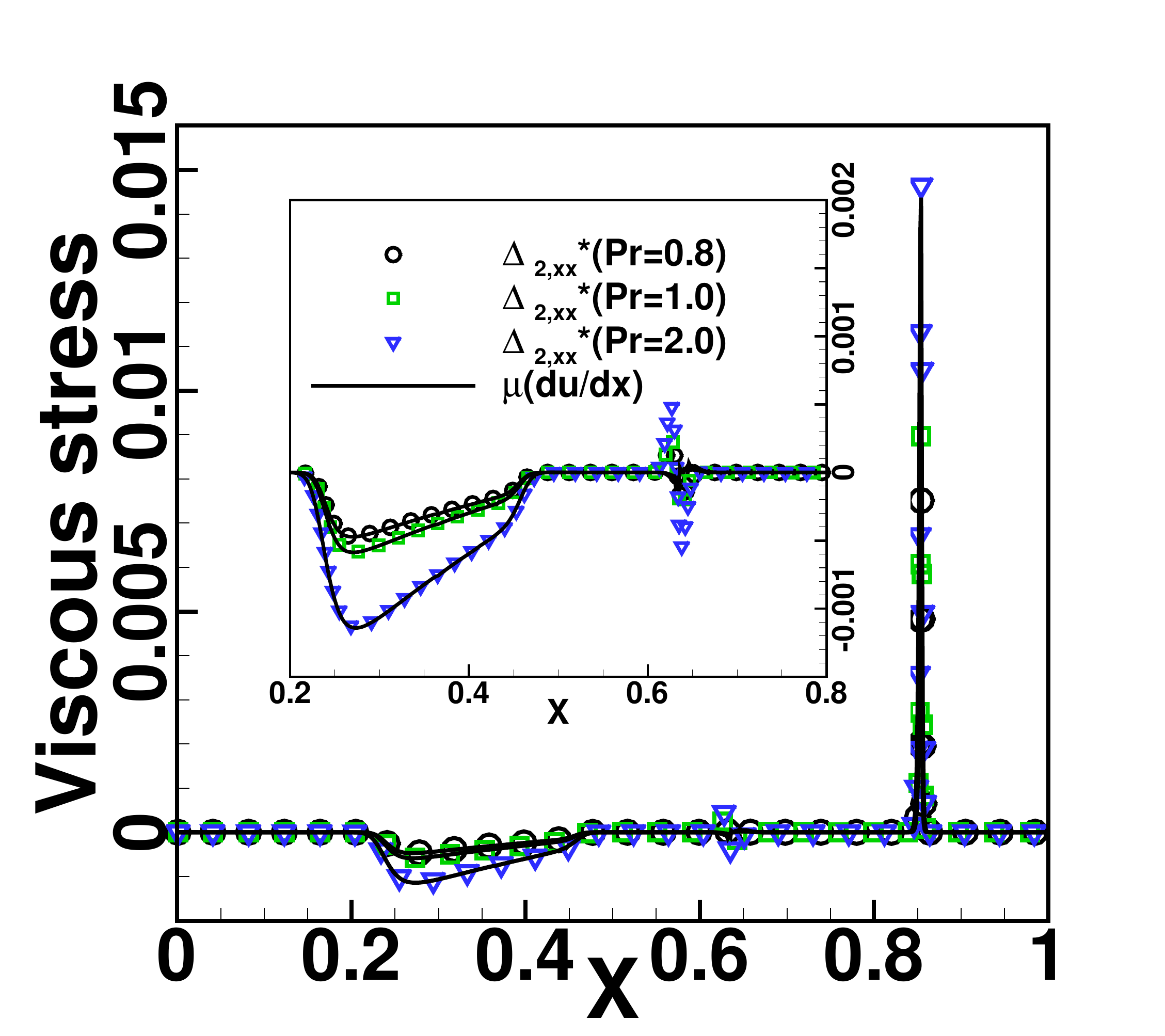}
\end{center}
\caption{Profiles of viscous stress $\Delta_{2,xx}^{S*}$ of the Sod' shock tube, at $t=0.18$. The lines indicate Riemann solutions, and the simulation results are denoted by circles, squares, and triangles, corresponding to Prandtl number $\Pr$=0.8, 1.0, and 2.0, respectively. The enlarged view is the profiles of $\Delta_{2,xx}^{S*}$ from 0.2 to 0.8 in $x$ axis.} \label{fig6}
\end{figure}

\subsubsubsection{The viscous stress of the system}
Here we give a method in calculating the viscous stress of the system. Figure \ref{fig6} shows the comparisons of viscous stress $\Delta_{2,xx}^{S*}$ between simulation results and analytical solutions of various evolutionary processes, corresponding to Prandtl number $\Pr$ = 0.8, 1.0, and 2.0, respectively. The symbol ``S'' represents fluid system, and $\Delta_{2,xx}^{S*}=\Delta_{2,xx}^{A*}+\Delta_{2,xx}^{B*}$. It means the viscous stress of the physical system is equal to the sum of the viscous stress of components A and B.
A small oscillation around contact discontinuity is captured by DBM simulations, which can not be provided by analytical solutions. The enlarged view, from 0.2 to 0.8 in $x$ axis, indicates the approximations between simulation results and analytical results. Moreover, the values of $\Delta_{2,xx}^{S*}$ with $\Pr = 2.0$ are larger than the cases of $\Pr=0.8$ and $\Pr=1.0$, indicating a farther distance to equilibrium state. Because of the larger Prandtl number, the larger values of viscous stress, which makes the system further away from equilibrium state. 

\subsubsection{Lax's shock tube}

\begin{figure}
\begin{center}
\includegraphics[width=0.5\textwidth]{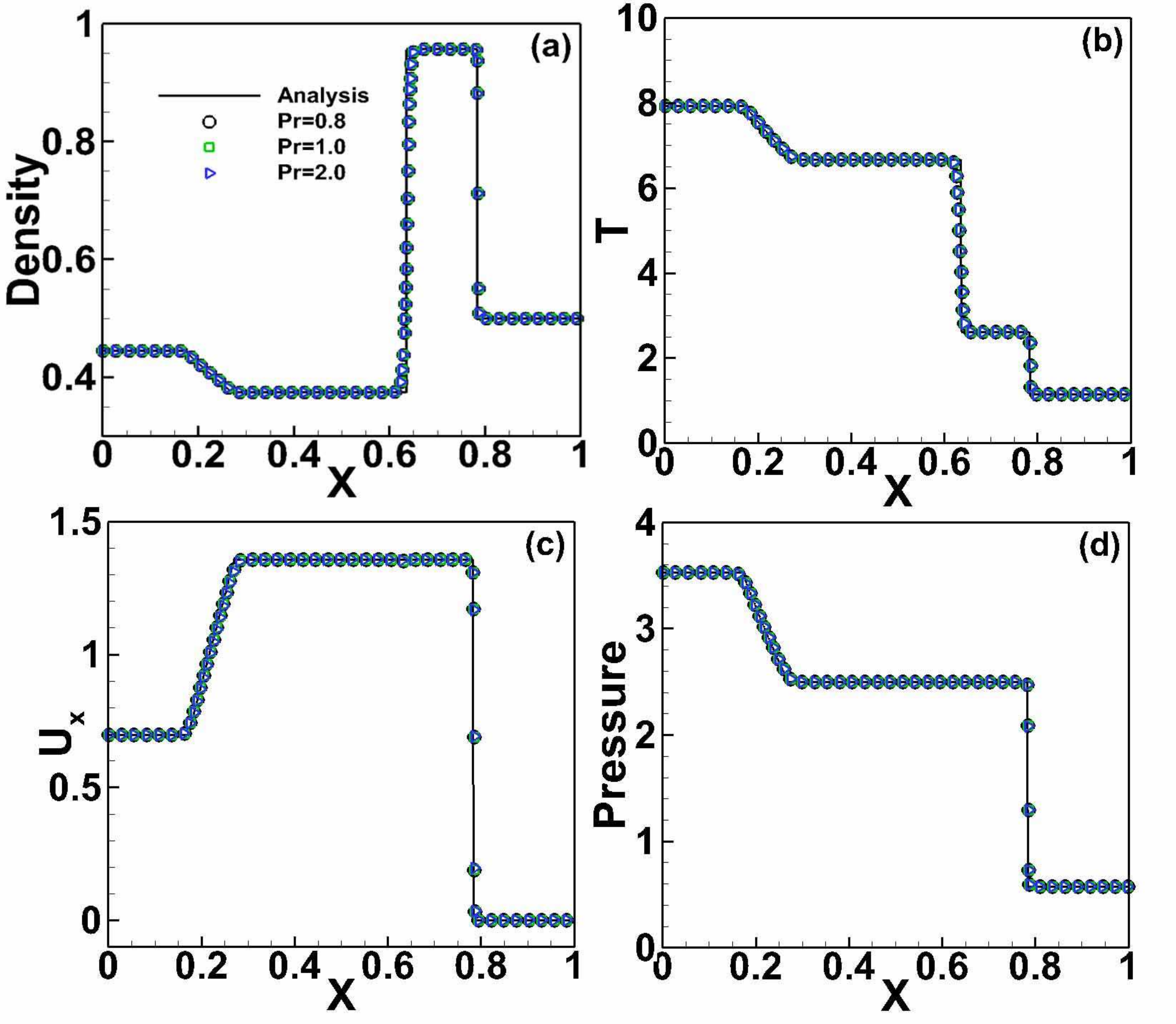}
\end{center}
\caption{Profiles of density (a), temperature (b), $U_{x}$ (c) and pressure (d) of the Lax's shock tube, at $t=0.1$. The lines indicate Riemann solutions, and the simulation results are denoted by circles, squares, and triangles, corresponding to Prandtl number $\Pr$ = 0.8, 1.0, and 2.0, respectively.} \label{fig7}
\end{figure}

The initial conditions of other quantities are $c=1,7$, $m^{A}=m^{B}=1$, $\tau^{A}=\tau^{B}=1\times 10^{-5}$, $I^{A}=I^{B}=0(\gamma^{A}=\gamma^{B}=2.0)$, $\Delta t=1\times10^{-5}$, $\Delta x=\Delta y=10^{-3}$,$\eta^{A}=\eta^{B}=0.0$.
Shown in Fig. \ref{fig7} are the comparisons between simulation results and Riemann solutions of density, temperature, velocity, and pressure profiles at $t = 0.1$, with the coefficient $b$ = -0.25, 0.0, and 0.5, respectively (corresponding to Prandtl number $\Pr$ = 0.8, 1.0, and 2.0, respectively). It is clear that a left-propagating rarefaction wave and the right-propagating shock wave with $\tt{Ma}$=1.87879 are captured by DBM. The simulation results of the shock wave rear are $(\rho,p)_{simulation} = (0.95748,2.49722)$. They are consistent with analytical solutions $(\rho,p)_{analysis} = (0.95749,2.49702)$ that calculated by Eqs. (\ref{Eq:DDBM-RH2}) and (\ref{Eq:DDBM-RH3}), indicating the ability of capturing 1-dimensional shock front accurately.

\subsubsection{Sjogreen's problem}

\begin{figure}
\begin{center}
\includegraphics[width=0.5\textwidth]{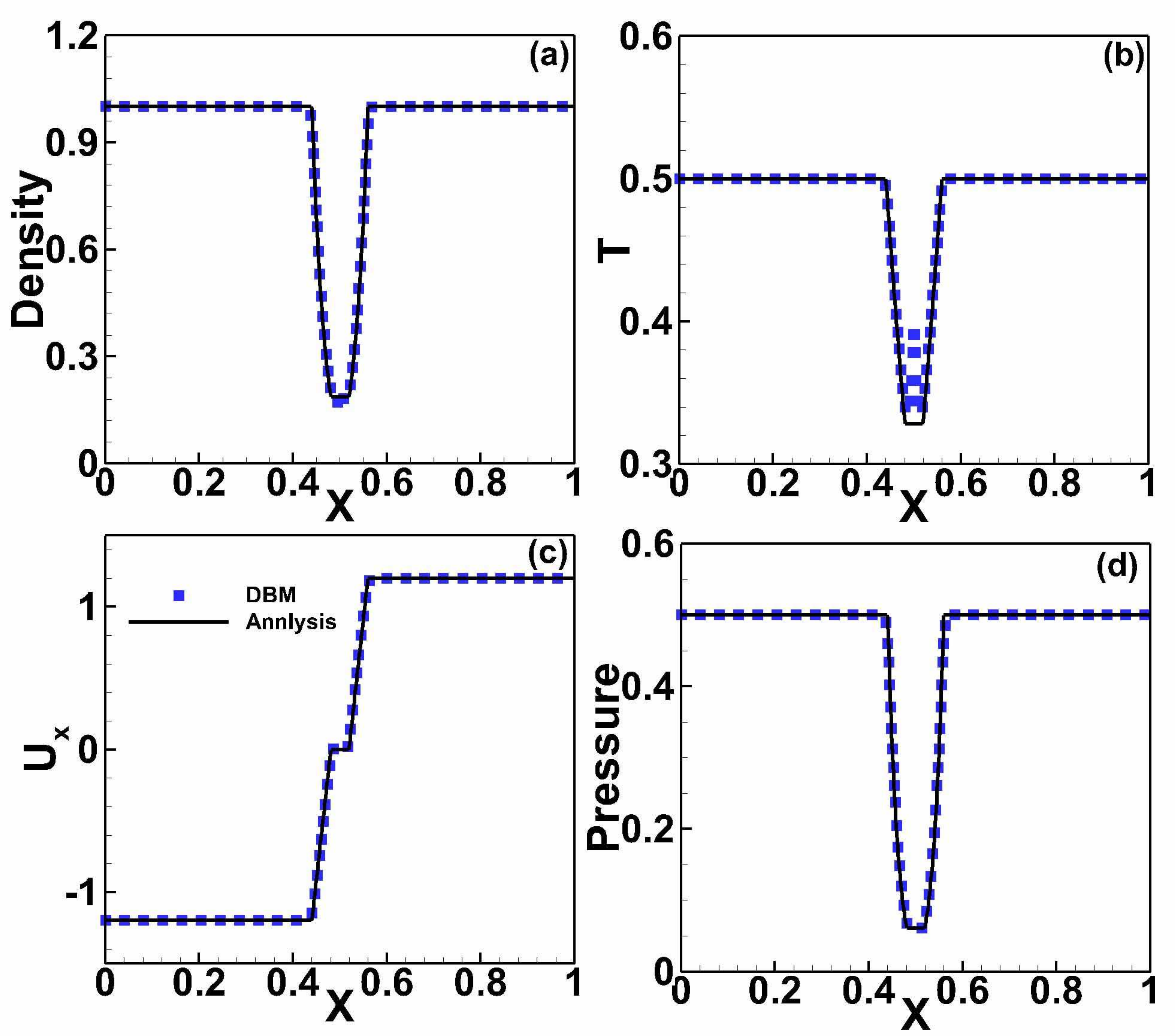}
\end{center}
\caption{Profiles of density (a), temperature (b), $U_{x}$ (c) and pressure (d) of the Sjogreen's problem at $t=0.03$ with Prandtl number $\Pr$ = 1.0. The lines indicate Riemann solutions, and the simulation results are denoted by blue squares.} \label{fig8}
\end{figure}

The initial conditions of other quantities are $c=0.8$, $m^{A}=m^{B}=1$, $\tau^{A}=\tau^{B}=2\times 10^{-5}$, $I^{A}=I^{B}=6(\gamma^{A}=\gamma^{B}=1.25)$, $\Delta t=1\times10^{-5}$, $\Delta x=\Delta y=10^{-3}$,$\eta^{A}=\eta^{B}=15.0$.
Shown in Fig. \ref{fig8} are the comparisons between simulation results and Riemann solutions of density, temperature, velocity, and pressure profiles at $t = 0.03$, with the coefficient $b$ = 0.0 (corresponding to Prandtl number $\Pr$ = 1.0). It is clear that a left-propagating rarefaction wave and a right-propagating rarefaction wave are captured by DBM. We also find the well agreement between DBM results and analytical solutions.

\subsubsection{The collision of two strong shock waves}

\begin{figure}
\begin{center}
\includegraphics[width=0.5\textwidth]{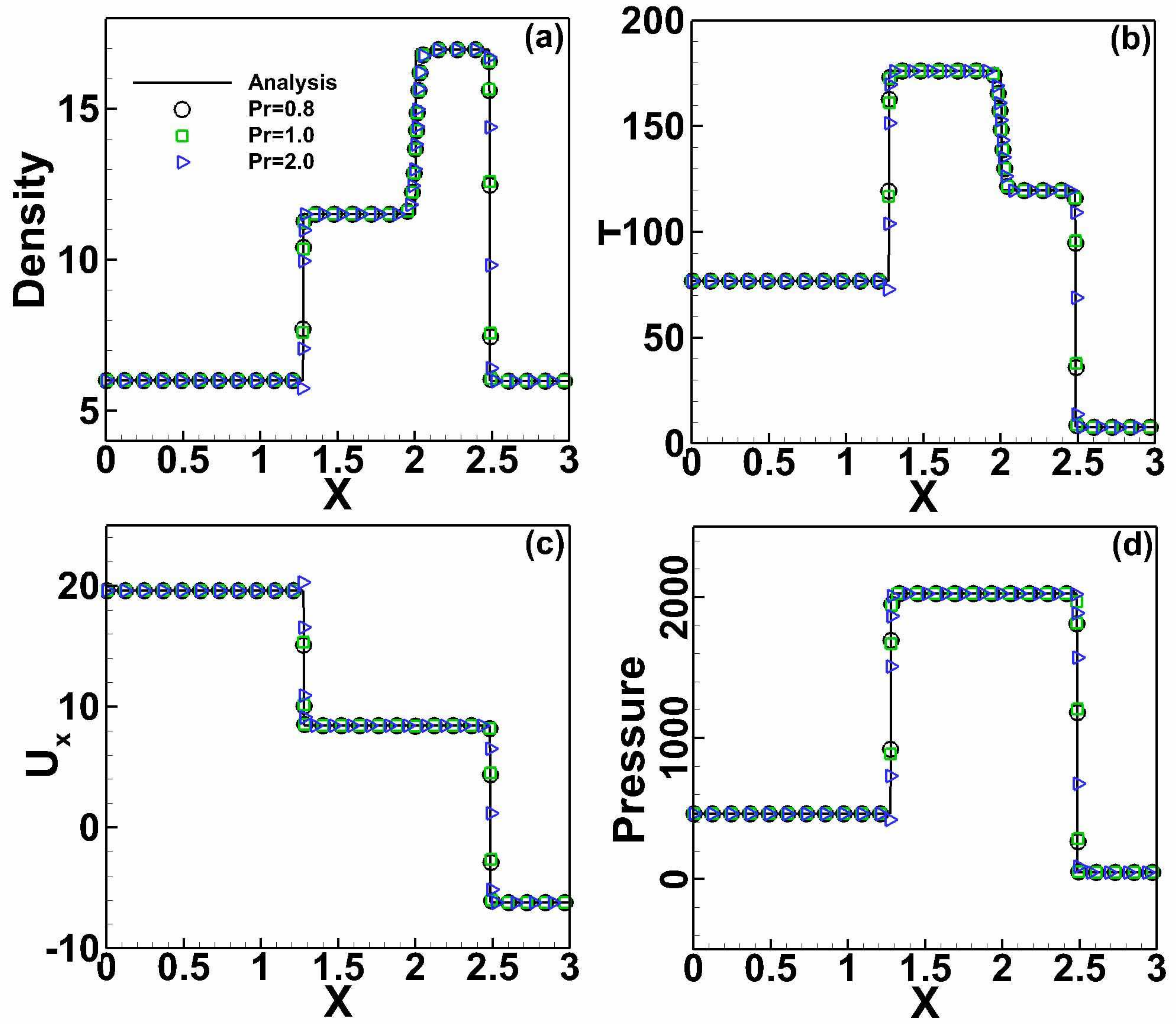}
\end{center}
\caption{Profiles of density(a), temperature(b), $U_{x}$(c) and pressure(d) of the collision of two strong shock waves, at $t=0.06$. The lines indicate Riemann solutions, and the simulations results are denoted by circles, squares, and triangles, corresponding to Prandtl number $\Pr$ = 0.8, 1.0, and 2.0, respectively.} \label{fig9}
\end{figure}

The initial conditions of other quantities are $c=8.0$, $m^{A}=m^{B}=1$, $\tau^{A}=\tau^{B}=2\times 10^{-5}$, $I^{A}=I^{B}=0(\gamma^{A}=\gamma^{B}=2.0)$, $\Delta t=1\times10^{-5}$, $\Delta x=\Delta y=3\times 10^{-3}$,$\eta^{A}=\eta^{B}=0$.
Shown in Fig. \ref{fig9} are the comparisons between simulation results and Riemann solutions of density, temperature, velocity, pressure profiles at $t = 0.06$, with the coefficient $b$ = -0.25, 0.0, and 0.5, respectively (corresponding to Prandtl number $\Pr$ = 0.8, 1.0, and 2.0, respectively). We find a good agreement between DBM results and analytical solutions. And, it is clear that a slow right-propagating shock wave with $\tt{Ma}$=1.88343 and a fast right-propagating shock wave with $\tt{Ma}$=5.76315 are captured by DBM. The simulation results of the left shock wave rear are $(\rho,p)_{simulation} = (11.50887,2026.27941)$, which are consistent with analytical solutions $(\rho,p)_{analysis} = (11.50891,2026.27969)$ according to Eqs. (\ref{Eq:DDBM-RH2}) and (\ref{Eq:DDBM-RH3}). The simulation results of the right shock wave rear are $(\rho,p)_{simulation} = (16.95647,2026.25441)$, which are consistent with analytical solutions $(\rho,p)_{analysis} = (16.95623,2025.96231)$ according to Eqs. (\ref{Eq:DDBM-RH2}) and (\ref{Eq:DDBM-RH3}).

\subsubsection{The Sod' shock tube of two components with different particle masses}

\begin{figure}
\begin{center}
\includegraphics[width=0.5\textwidth]{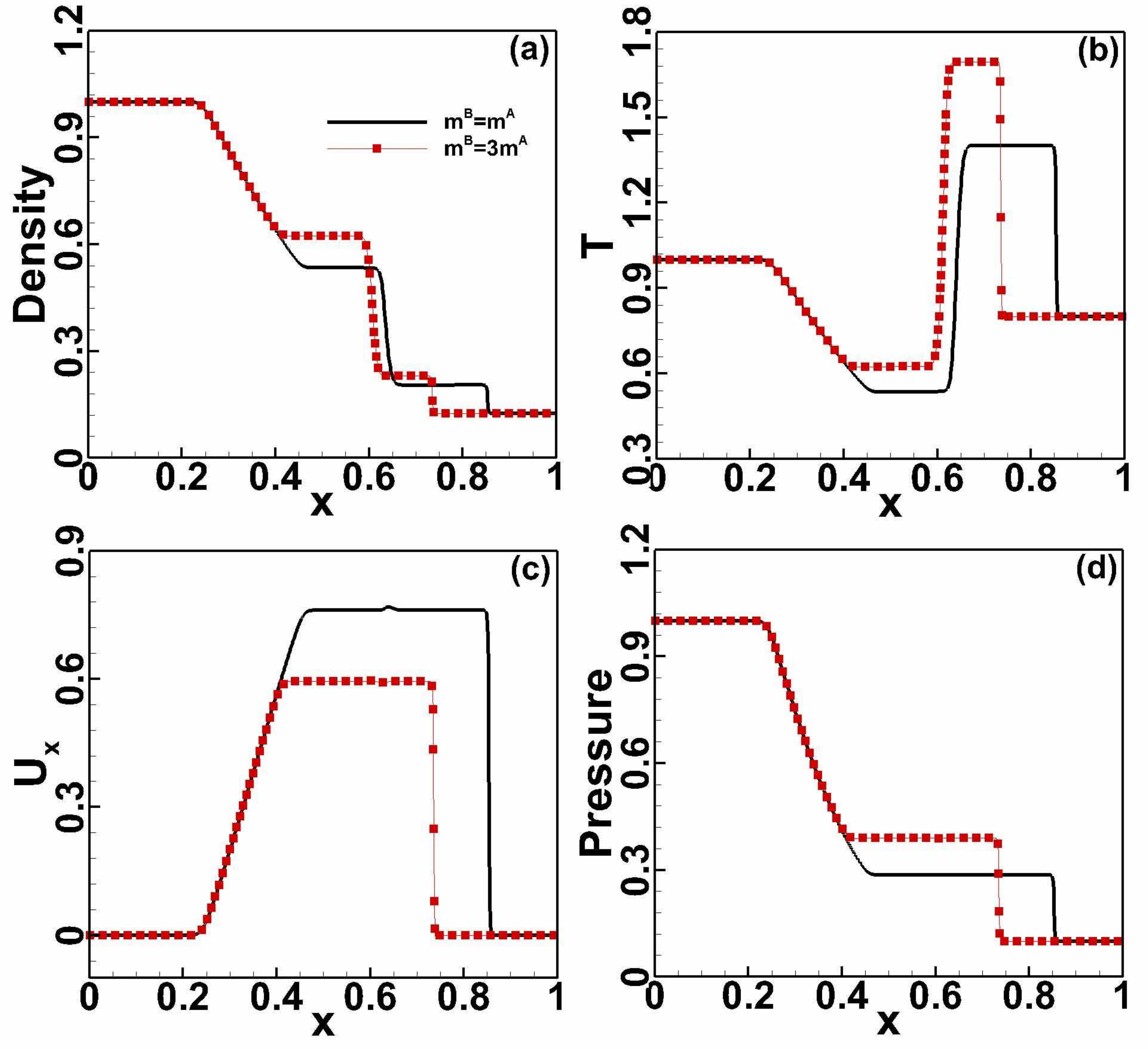}
\end{center}
\caption{Profiles of density(a), temperature(b), $U_{x}$(c), and pressure(d) at $t=0.18$ with $\Pr=1.0$, respectively. The black lines and red lines with squares represent simulation results with $m^{B}=m^{A}$ and $m^{B}=3m^{A}$, respectively.} \label{fig10}
\end{figure}

The particle mass difference is not taken into account in the simple Riemann analytical solution. Besides, a single-fluid DBM can not capture effectively the behaviors of two-fluid system with different particle masses. Actually, the difference of particle masses of two components makes sense to the evolution. In the following, we use our two-fluid DBM to simulate a Sod's shock tube of two components with different particle masses, and investigate the effects of particle mass difference.
Except for particle masses $m^{B}=3m^{A}=3$, other parameters are the same as shown in section \ref{Sod' shock tube}.
Figure \ref{fig10} are the profiles of density(a), temperature(b), $U_{x}$(c), and pressure(d) at $t=0.18$ with $\Pr=1.0$, respectively. The black lines and red lines with squares represent simulation results with $m^{B}=m^{A}$ and $m^{B}=3m^{A}$, respectively. In the case with $m^{B}=3m^{A}$, the contact discontinuity interface and shock wave interface move slower than the case with $m^{B}=m^{A}$. Because the smaller $m^{A}$ of component A and the larger $m^{B}$ of component B correspond to ``light'' fluid and ``heavy'' fluid, respectively. Thus, compared to the case with $m^{B}=m^{A}$, the ``light'' fluid cause these slower interfaces in this case with $m^{B}=3m^{A}$.

\subsection{Kelvin-Helmholtz instability}

The Kelvin-Helmholtz instability(KHI) is an efficient and significant mechanism for turbulence and mixing of fluids in ICF. It occurs at a perturbed interface between two fluids or two parts of the same fluid with different densities and tangential velocities. At present, the single-fluid DBM has brought meaningful progress in KHI investigation\cite{2019Gan-KHI,2011Gan-LBMKHI}. Instead of simulating KHI phenomenon with a single-fluid DBM, we here use the two-fluid DBM to investigate this typical two-dimensional complex flow, that enables us to simulate the KHI system which is composed of two different components. The initial condition is given as follows.
\[
\left\{
\begin{array}{l}
n^{\sigma}(x)=\frac{{n^{\sigma}_{L}+n^{\sigma}_{R}}}{2}-\frac{{n^{\sigma}_{L}-n^{\sigma}_{R}}}{2}\tanh (\frac{x-x_{0}+W \cos(k y)}{{D_{\rho }}}) \tt{,} \\
\mathbf{u}(x)=\frac{{\mathbf{u}_{L}+\mathbf{u}_{R}}}{2}-\frac{\mathbf{u}_{L}{-\mathbf{u}_{R}}}{2}\tanh (\frac{x-x_{0}+W \cos(k y)}{{D_{u}}}) \tt{,} \\
p(x)=p_{L}=p_{R} \tt{,}
\end{array}
\right.
\]
where $n^{\sigma}_{L}(n^{\sigma}_{R})$, $\mathbf{u}^{\sigma}_{L}(\mathbf{u}^{\sigma}_{R})$, $p_{L}(p_{R})$ are the particle number density, the velocity of fluid system, and the pressure of component $\sigma$ near the left(right) boundary, respectively; $D_{\rho}(D_{u})$ is the width of density(velocity) transition layer; $x_{0}$ is the average $x$ position of material interface; $W$ is the perturbation amplitude in initial condition; $k$ is the perturbation wave number. Thus, the two components have the same velocity and temperature at the same place. Furthermore, inflow/outflow(zero gradient) boundary conditions and the periodic boundary conditions are adopted in $x$ direction and $y$ direction, respectively. The parameters are chosen as $n^{A}_{L}=0.8$, $n^{A}_{R}=0.2$, $n^{B}_{L}=0.2$, $n^{B}_{R}=0.8$, $\mathbf{u}_{L}=u_{L} \mathbf{e}_{y}$, $\mathbf{u}_{R}=u_{R}\mathbf{e}_{y}$, $u_{L}=0.5$, $u_{R}=-0.5$, $D_{u}=D_{\rho}=L_{x}/80$, $W=L_{x}/100$, $x_{0}=L_{x}/2$, $k_{\lambda}=2\pi/L_{y}$, $\tau^{A}=\tau^{B}=1\times10^{-5}$, $\Delta t=5\times 10^{-6}$, $L_{x}=L_{y}=0.2$, $m_{A}=m_{B}=1.0$, $p_{L}=p_{R}=1.0$, $I^{A}=I^{B}=0$, $\eta^{A}=\eta^{B}=0$, $b=0.0$.

\begin{figure}
\begin{center}
\includegraphics[width=0.45\textwidth]{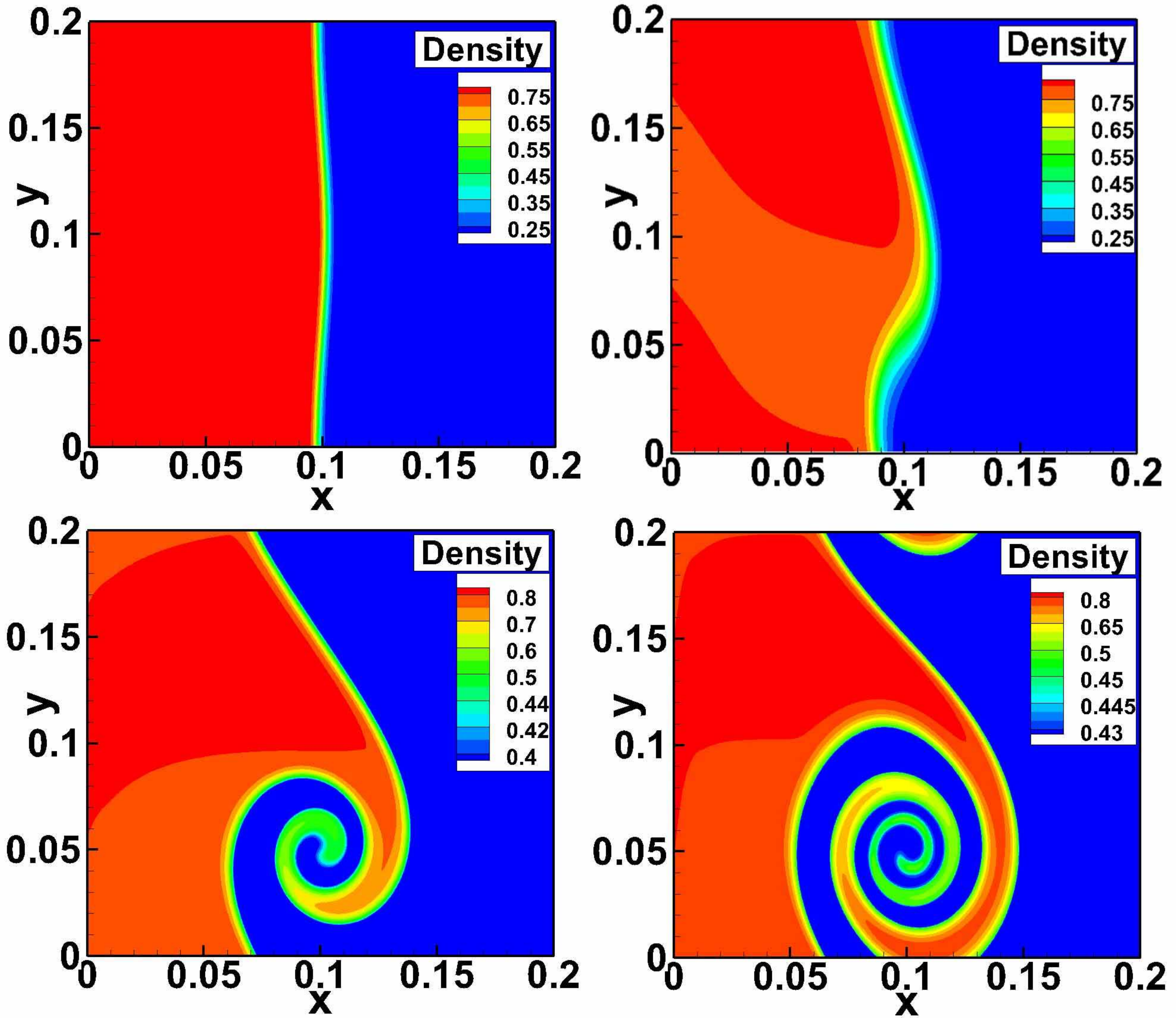}
\end{center}
\caption{ Density contours of component A in the evolutions of KHI at various times, $t$ = 0.0, 0.2, 0.4, and 0.6, respectively, with $\Pr$ = 1.0. The color from blue to red indicates the increase of density.} \label{fig11}
\end{figure}

\begin{figure}
\includegraphics[width=0.45\textwidth]{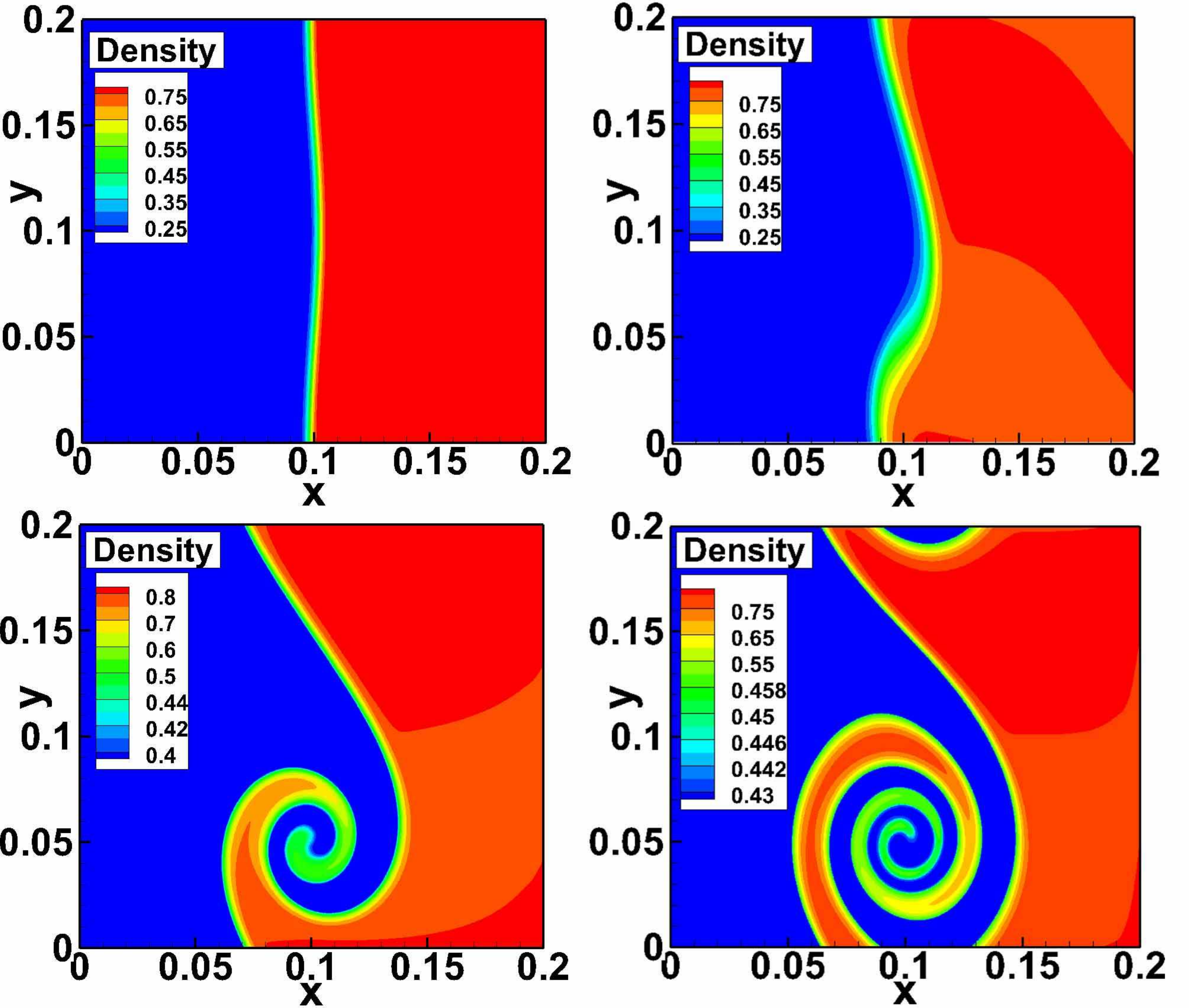}
\caption{ Density contours of component B in the evolutions of KHI at various times, $t$ = 0.0, 0.2, 0.4, and 0.6, respectively, with $\Pr$ = 1.0. The color from blue to red indicates the increase of density.} \label{fig12}
\end{figure}

This simulation has good numerical stability and low computational costs. The computational facility used here is a personal computer with Intel(R) Core(TM) i5-9400F CPU @2.90GHz and RAM 16.00GB, with 109991.047s operation time(in the case with grid $400\times400$). Figure \ref{fig11} shows the density contours of component A in the evolution of KHI at four different times, with $\Pr=1.0$. It is evident that the interface is distorted by pressure difference at $t$ = 0.2. After the initial linear growth stage, a roll-up vortex formulates around the interface at $t$ = 0.4. Then, at $t$ = 0.6, a larger vortex is observed in the density field. Similarly, the density contours of component B in the evolution of KHI are shown in Fig. \ref{fig12}.

\subsubsection{Grid convergence test}
\begin{figure}
\begin{center}
\includegraphics[width=0.4\textwidth]{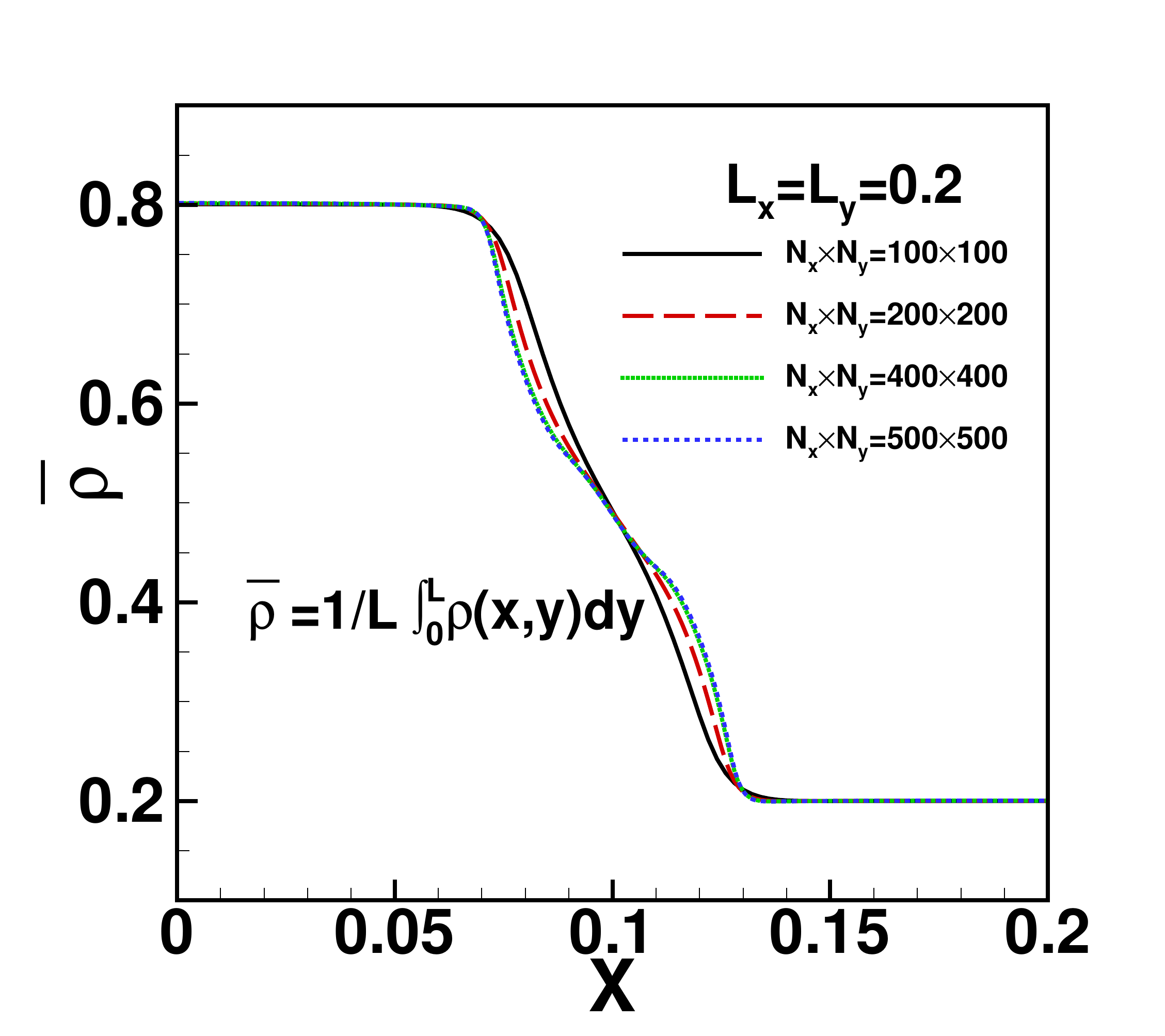}
\end{center}
\caption{Grid convergence test of KHI simulation: Profiles of $\overline \rho^{A}$ against the $x$ axis at $t=0.3$, with four different mesh grids. The black line, red line, green line, and blue line corresponding to mesh grids $N_{x}\times N_{y}=100\times100$, $200\times200$, $400\times400$, and $500\times500$, respectively.} \label{fig13}
\end{figure}

Grid convergence is important in numerical simulations. To verify the validity of the simulations, we carry out the grid convergence test by using various grids: $N_{x}\times N_{y}=100\times100$, $200\times200$, $400\times400$, and $500\times500$. Shown in Fig. \ref{fig13} are the profiles of $\overline \rho^{A}$ against the $x$ axis at $t=0.3$, with four different mesh grids. The black line, red line, green line, and blue line corresponding to mesh grids $N_{x}\times N_{y}=100\times100$, $200\times200$, $400\times400$, and $500\times500$, respectively. The averaged density $\overline \rho^{A}$, which is defined as $\overline \rho^{A}(x)=\frac{1}{L}\int^{L}_{0}\rho^{A}(x,y)dy$, is important to quantitatively describe the characteristics of the vortex of the mixing layer\cite{2011Gan-LBMKHI}. As we can see, the profiles $\overline \rho^{A}$ with grids $400\times400$ and $500\times500$ are almost coincide. Taking account of both accuracy and computational costs, we carry out this simulation on the grid $400\times400$.

\subsubsection{Comparison with analytical solution}
\begin{figure}
\begin{center}
\includegraphics[width=0.4\textwidth]{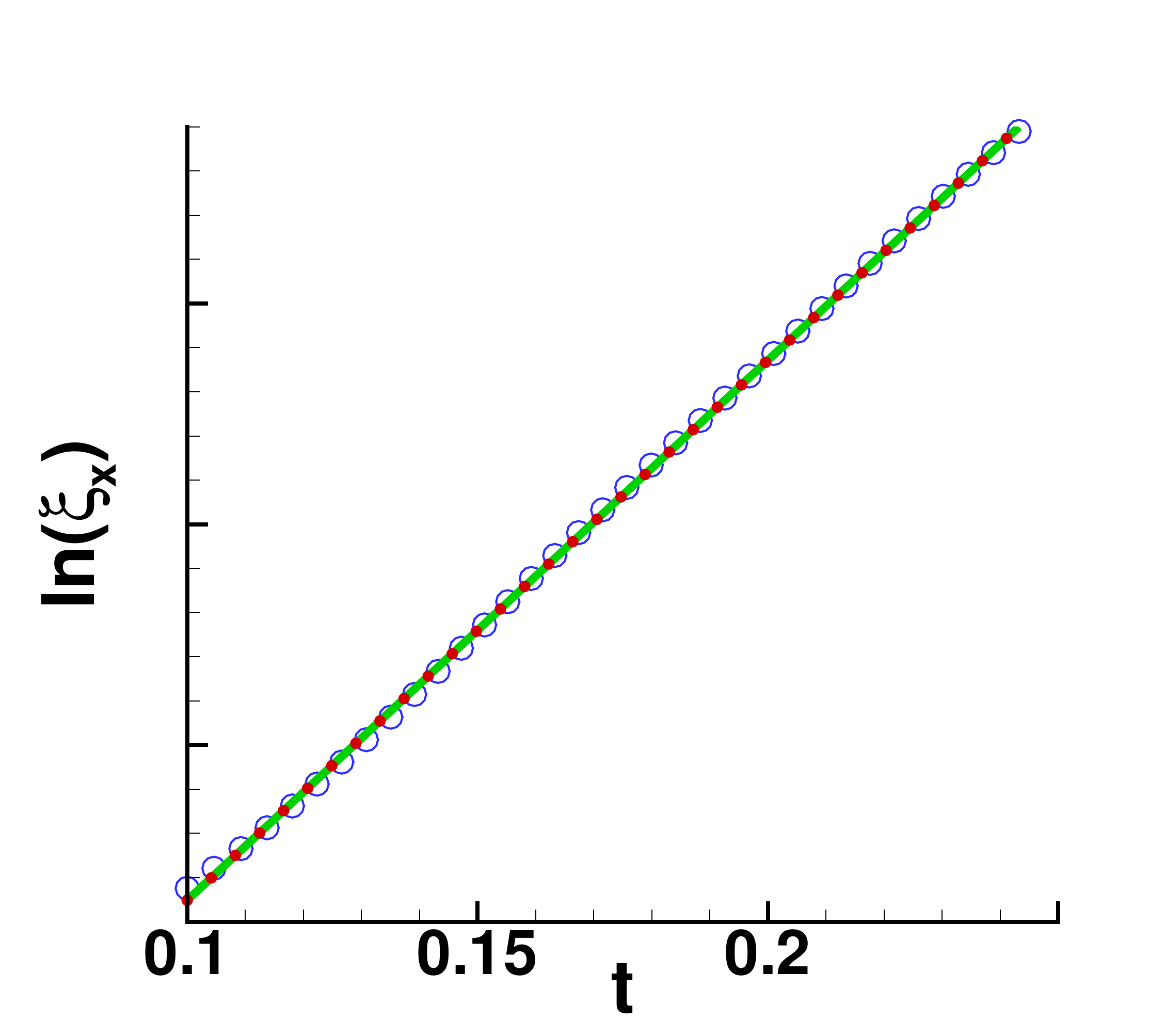}
\end{center}
\caption{The profile of $\ln(\xi_{x})$ within the linear stage (0.1<$t$<0.25) of the KHI with $\Pr=2.0$. The blue circles represent DBM results, the continuous green line denotes the fitting function, and the red dots is for the analytical solution.} \label{fig14}
\end{figure}

Here we quantitatively compare the DBM results with the analytical solution. In Fig. \ref{fig14}, we show the perturbed peak kinetic energy $\xi_{x}=\frac{1}{2}\rho^{A} (u_{x}^{A})^{2}$ versus time $t$ in the evolution of KHI with $\Pr=2.0$. The profile of $\ln(\xi_{x})$ within the linear stage (0.1<$t$<0.25) of the KHI is plotted. The blue circles represent DBM results, the continuous line denotes the fitting function $F(t)=-9.14803+24.4539t$, and the red dots is for the analytical solution $F(t)=-9.14803+2\dot{A}t$ where $\dot{A}=12.39925$ is half linear growth rate of $\xi_{x}$\cite{2010Local,2011Gan-LBMKHI,2016Lin-CNF}. The growth rate can be calculated by Eq. (18) in Ref.\cite{2010Combined} . The relative difference between DBM results and analytical solution is $-1.4\%$.

\subsubsection{The influence of Pr number in KHI}
\begin{figure*}
\begin{center}
\includegraphics[width=0.7\textwidth]{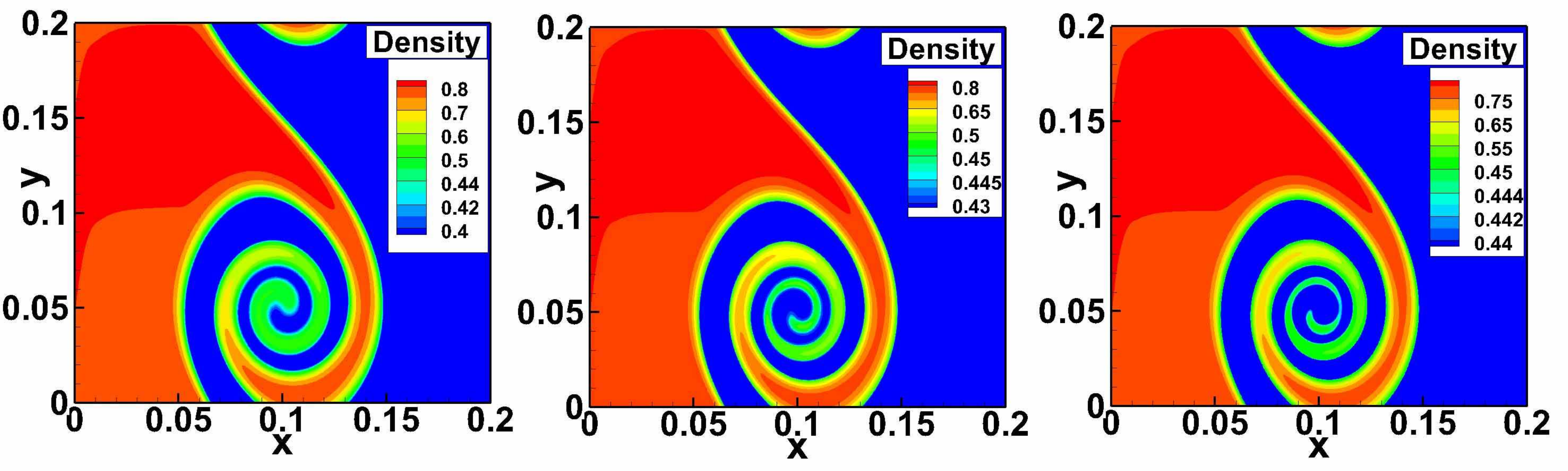}
\end{center}
\caption{Density contours of component A at $t=0.6$, with Prandtl number $\Pr$ = 2.0, 1.0, and 0.8, respectively. The color from blue to red indicates the increase of density.} \label{fig15}
\end{figure*}

\begin{figure*}
\begin{center}
\includegraphics[width=0.7\textwidth]{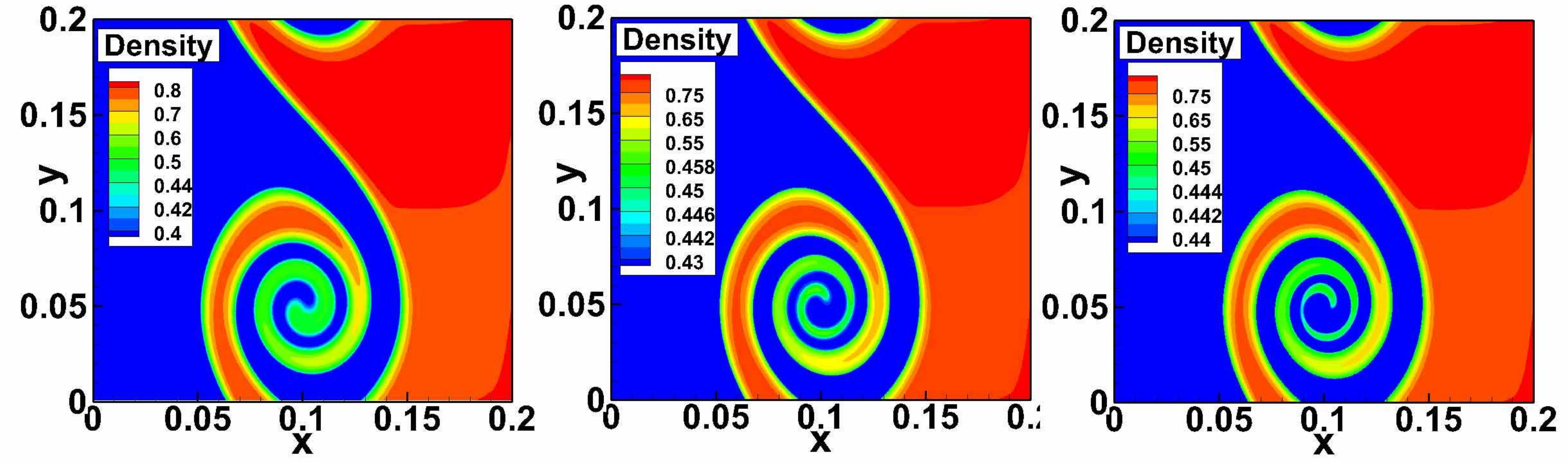}
\end{center}
\caption{Density contours of component B at $t=0.6$, with Prandtl number $\Pr$ = 2.0, 1.0, and 0.8, respectively. The color from blue to red indicates the increase of density.} \label{fig16}
\end{figure*}

To investigate the influence of Prandtl number on the evolution of KHI, we conduct three runs with various Prandtl numbers, $\Pr$ = 2.0, 1.0, and 0.8, respectively. Figure \ref{fig15} shows the density contours of component A at $t=0.6$, with Prandtl numbers, $\Pr$ = 2.0, 1.0, and 0.8, respectively. It can be observed that a higher Prandtl number corresponds to a slower evolution. Because the higher the Prandtl number, the greater the viscosity of the fluid, which would impede evolution. Similar behavious of density contours of component B are shown in Fig. \ref{fig16}.

\subsubsection{The TNE behaviours on KHI}
\begin{figure}
\begin{center}
\includegraphics[width=0.45\textwidth]{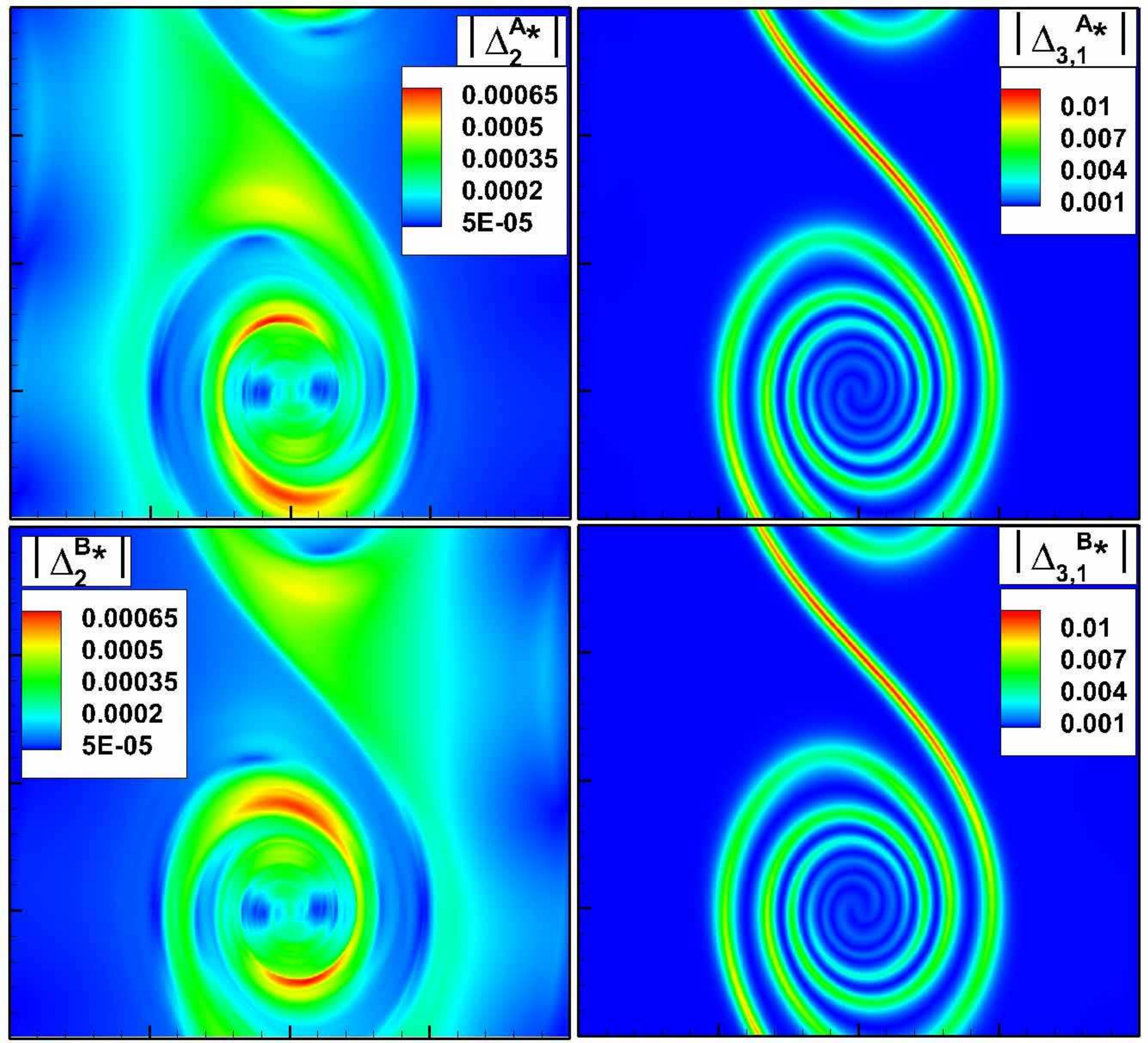}
\end{center}
\caption{$\left|\bm{\Delta}_{2}^{\sigma*}\right|$ and $\left|\bm{\Delta}_{3,1}^{\sigma*}\right|$ contours of component A and B at $t=0.6$, with Prandtl number $\Pr$ = 1.0, respectively.} \label{fig17}
\end{figure}

As mentioned above, DBM can supplement TNE information that is not available in the NS model. Besides, a two-fluid DBM can describe TNE behaviors of component A, component B, and physical system, respectively, which can not be achieved in a single-fluid DBM. Preliminarily, we study two kinds of TNE behaviors, $\left|\bm{\Delta}_{2}^{\sigma*}\right|$ and $\left|\bm{\Delta}_{3,1}^{\sigma*}\right|$, in the evolution of KHI. Shown in Fig. \ref{fig17} are the $\left|\bm{\Delta}_{2}^{\sigma*}\right|$ and $\left|\bm{\Delta}_{3,1}^{\sigma*}\right|$ contours of components A and B at $t=0.6$, with Prandtl number $\Pr$ = 1.0, respectively. We can see that the values of $\left|\bm{\Delta}_{2}^{A*}\right|$ and $\left|\bm{\Delta}_{2}^{B*}\right|$ are greater than zero around the vortex where the viscous stress is significant, while they are close to zero where far away from the interface. Meanwhile, the values of $\left|\bm{\Delta}_{3,1}^{A*}\right|$ and $\left|\bm{\Delta}_{3,1}^{B*}\right|$ are larger at the contact between two components while they approach zero where the interaction between the components is weak.

\begin{figure}
\begin{center}
\includegraphics[width=0.4\textwidth]{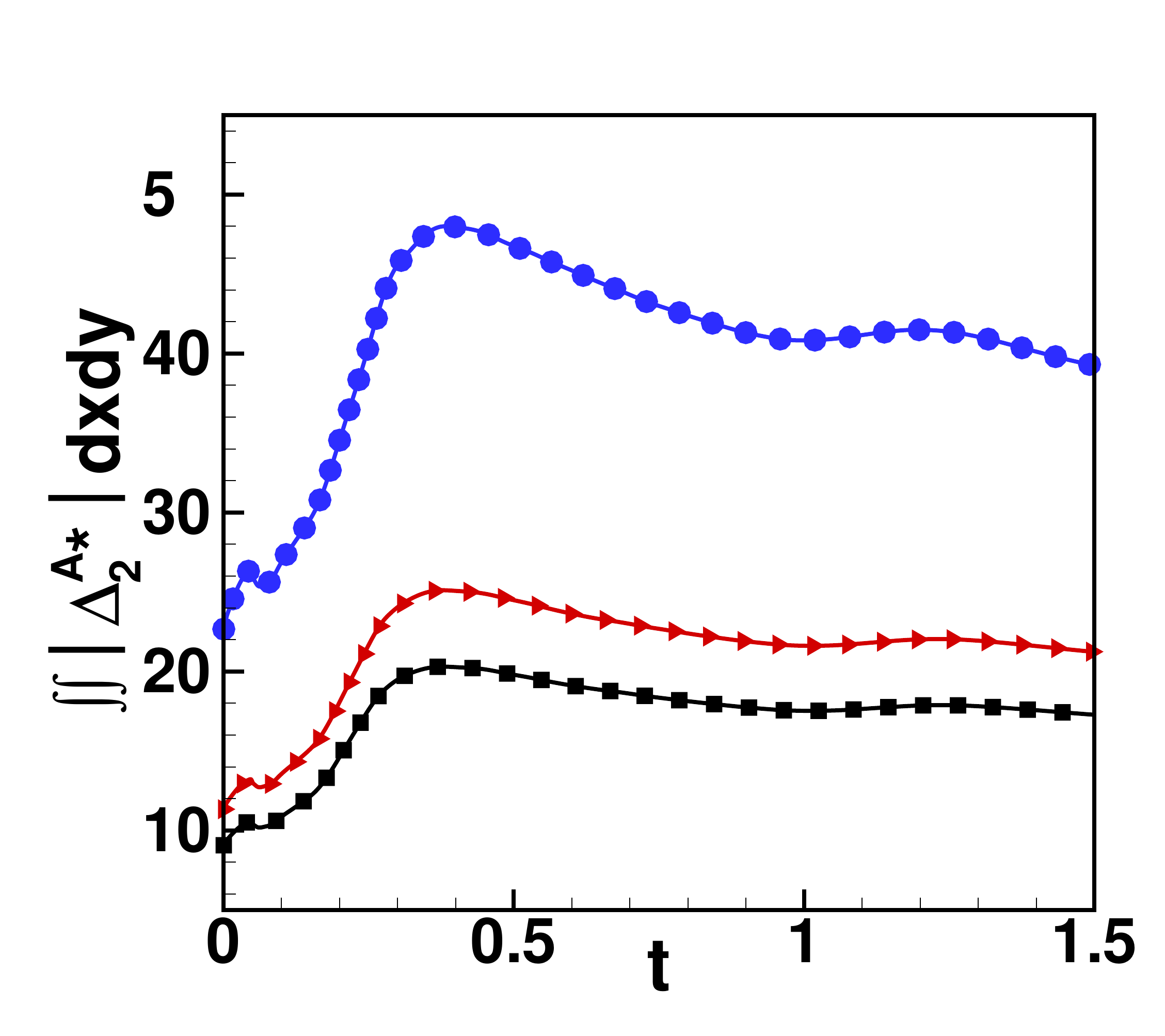}
\end{center}
\caption{Evolutions of $\int\int\left|\bm{\Delta}_{2}^{A*}\right|dxdy$ with various Prandtl numbers: $\Pr=0.8$(black line with squares), 1.0(red line with triangles), and 2.0(blue line with circles), respectively.} \label{fig18}
\end{figure}

To investigate the influence of the Prandtl number on global non-equilibrium effect(GNE) on the evolution of KHI, we conduct three runs with various Prandtl numbers, $\Pr$ = 0.8, 1.0, and 2.0, respectively. Plotted in Fig. \ref{fig18} are the evolutions of GNE of $\int\int\left|\bm{\Delta}_{2}^{A*}\right|dxdy$ with various Prandtl numbers, where the integral is extended over all physical space $L_{x}\times L_{y}$. The lines with squares, triangles, and circles corresponding to $\Pr$ = 0.8, 1.0, and 2.0, respectively. Actually, $\int\int\left|\bm{\Delta}_{2}^{A*}\right|dxdy$ represents the global strength of the viscosity of component A. It is evident that the GNE become stronger for larger Prandtl number and shows alternate increase-decline trends. Physically, there are competitive mechanisms in the evolution of GNE. The GNE is associated with the lengthened and widened interface, which would strengthen and weaken the GNE, respectively. The GNE of component B has the similarly behaviours as component A, which is not be shown here.

\begin{figure}
\begin{center}
\includegraphics[width=0.4\textwidth]{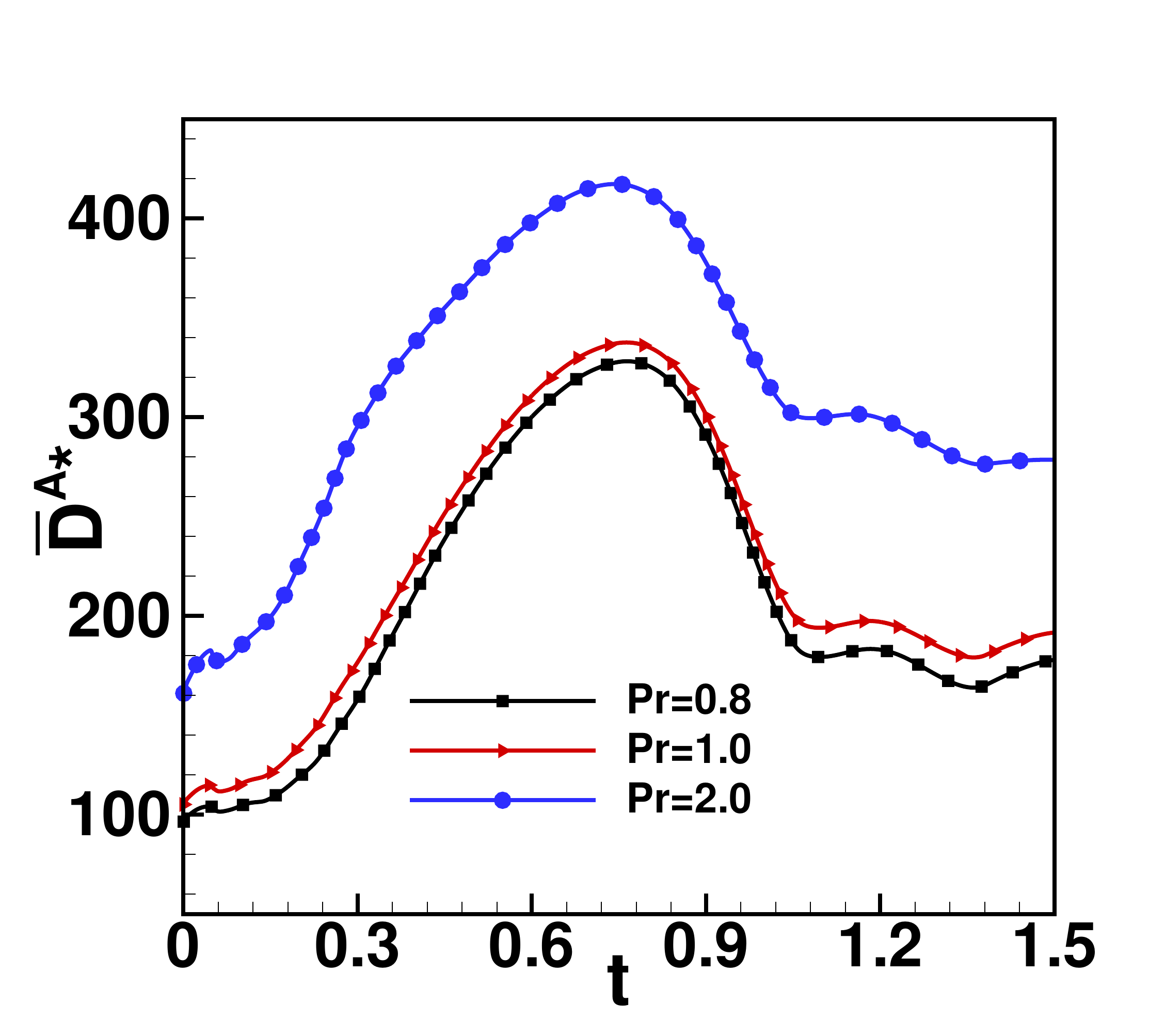}
\end{center}
\caption{The time evolution of the global average TNE strength $\overline{D}^{A*}$ with different Prandtl number: $\Pr=0.8$(black line with squares), 1.0(red line with triangles), and 2.0(blue line with circles), respectively.} \label{fig19}
\end{figure}

To investigate the influence of Prandtl number on TNE strength on the evolution of KHI, we conduct three runs with various Prandtl numbers, $\Pr$ = 0.8, 1.0, and 2.0, respectively.
Shown in Fig. \ref{fig19} are the evolutions of TNE strength $\overline{D}^{A*}$ with different Prandtl number. The profiles of global average TNE strength also show alternate increase-decline trends because of competition mechanisms in the evolution. Besides, the larger $\Pr$ number, the stronger TNE strength.
The TNE strength of component B has the similar behaviors with component A, which is not be shown here.

\subsection{Regular reflection of a shock wave}

The reflection of an oblique shock wave over a horizontal plane results in two types of wave configurations, regular reflection (RR) and Mach reflection (MR)\cite{2010Chen-MRT,2013Gan-LBGK}. Such a shock reflection problem is of great significance in both fundamental research and engineering applications. Such a supernova explosions in natural phenomena, hypersonic aircraft and ICF in engineering. Simply, the RR process can be seen as a single-fluid flow. In the following, we use two-fluid model to simulate this process, by setting one of the component $\rho=0$ initially. In other words, a two-fluid DBM can be reduced to a single-fluid DBM when neglecting the component differences. In this simulation, there is just component A in the flow filed, that is $(\rho,T,U_{x},U_{y})_{B}=0.0$. The coming shock(component A) has an angle of $25^{\circ}$, with Mach number 30. The computational domain is a rectangle with length of 3.0 and height of 1.0, which is divided into $300\times100$ rectangular grids. Other parameters are $b=0.0$, $m^{A}$=1.0, $\Delta t=1\times 10^{-5}$, $\tau^{A}=2\times 10^{-5}$, $I^{A}=-1.141262$, $c=18.0$, $\eta^{A}=12.0$. The boundary conditions are adopted a reflecting surface along the bottom boundary, outflow along the right boundary, and Dirichlet conditions on the left and upper boundary, respectively.
\[
\left\{
\begin{array}{l}
(\rho,T,U_{x},U_{y})_{0,y,t}=(1.0,1/3.329,0.0,0.0)                \tt{,} \\
(\rho,T,U_{x},U_{y})_{x,1.5,t}=(1.84886,40.0803,27.5399,-5.27567) \tt{.}
\end{array}
\right.
\]
\begin{figure}
\begin{center}
\includegraphics[width=0.5\textwidth]{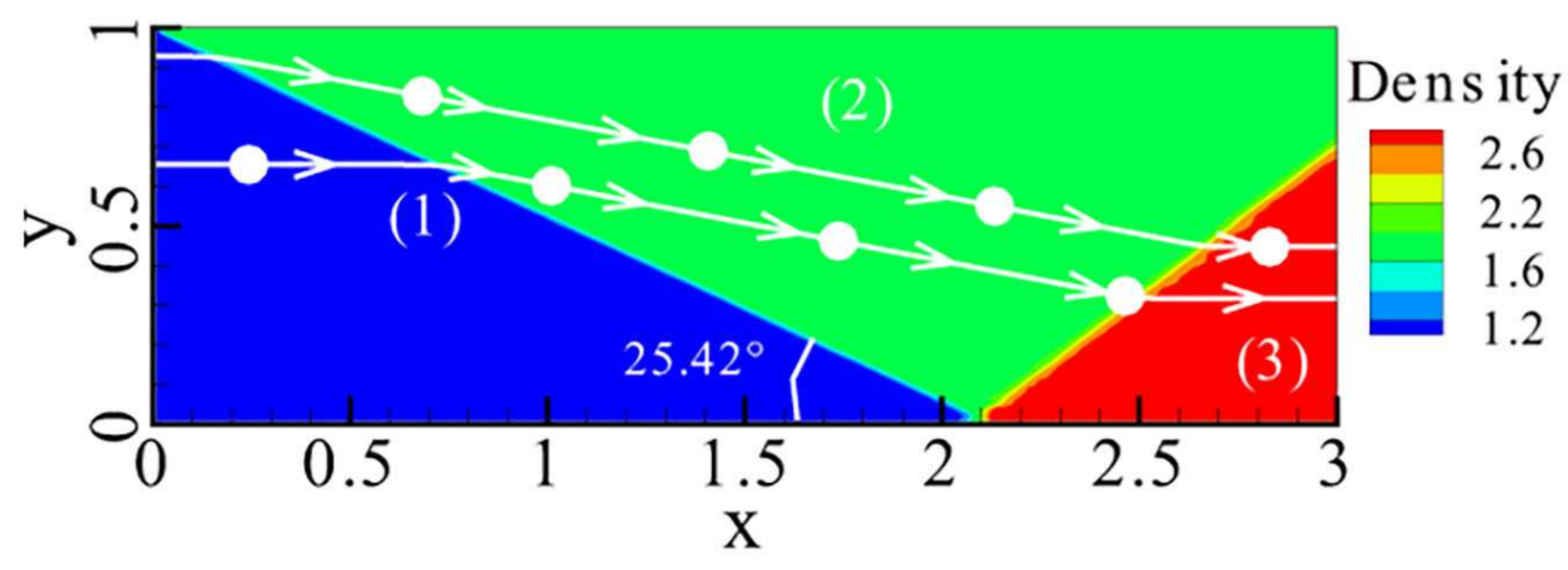}
\end{center}
\caption{Density contour along the $x$ direction of RR on a wall with $\Pr=1.0$ at $t=6.0$. The white lines and white spots represent streamline and virtual particles, respectively. The color from blue to red indicates the increase of density.} \label{fig20}
\end{figure}

\begin{figure}
\begin{center}
\includegraphics[width=0.4\textwidth]{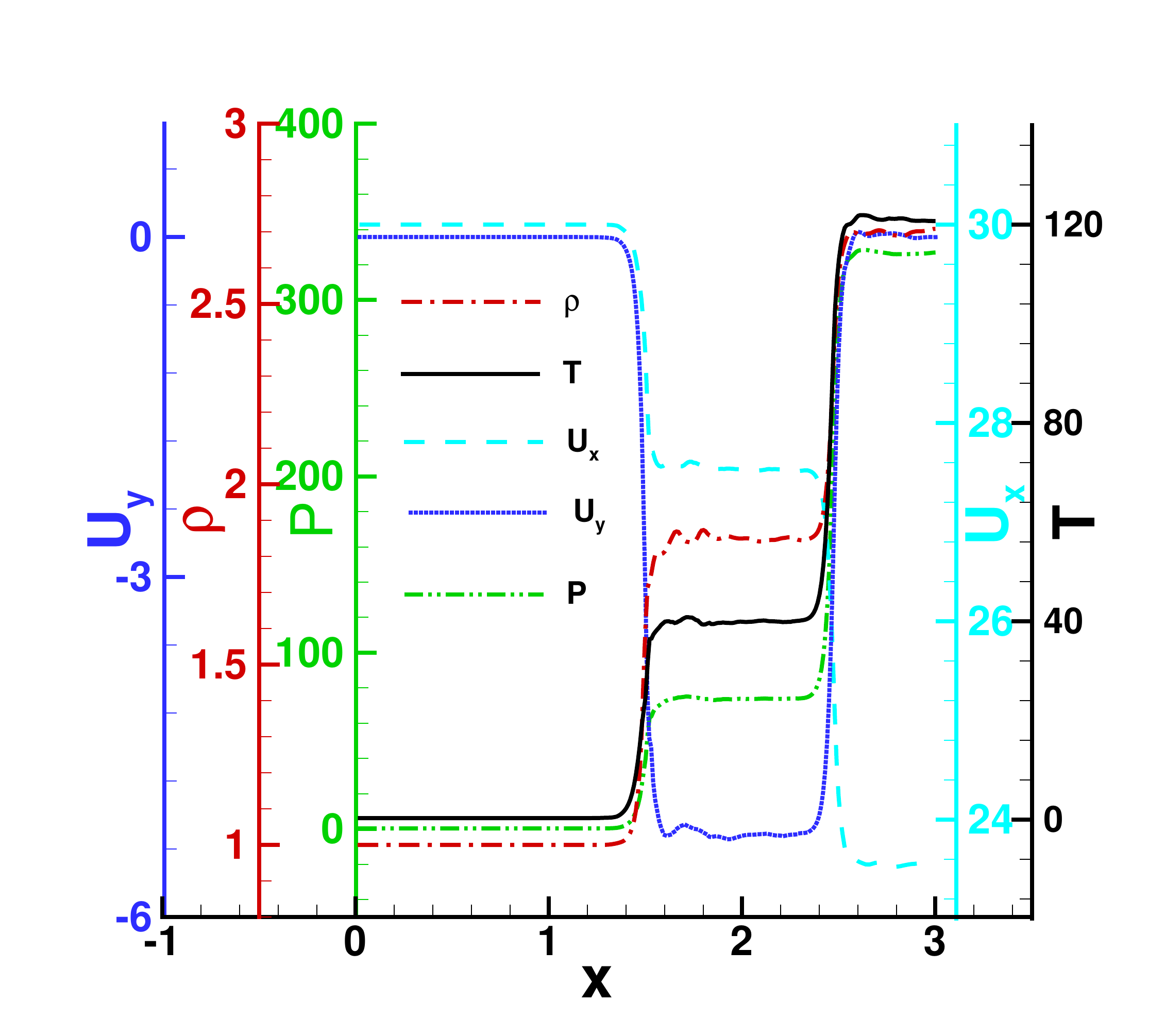}
\end{center}
\caption{Profiles of density(red line), temperature(black line), $U_{x}$(wathet line), $U_{y}$(dark blue line), and pressure(green line) at $N_{y}=30$ along $x$ direction with $\Pr=1.0$, at $t=6.0$.} \label{fig21}
\end{figure}

\subsubsection{Comparison with analytical solution}
Shown in Fig. \ref{fig20} is the density contour along the $x$ direction of RR on a wall. The white lines indicate the motion traces of virtual particles. Moreover, we can see the reflection angle $25.42^{\circ}$, which is approach the same with theoretical value of $25^{\circ}$. In addition, the shock wave divides the flow field into three parts, that the quantities of Part 1 and Part 2 or the quantities of Part 2 and Part 3 are both satisfied with Eqs. (\ref{Eq:DDBM-RH4}) and (\ref{Eq:DDBM-RH5})\cite{2015Xu-PRE}.
Shown in Fig. \ref{fig21} are the profiles of density, temperature, $U_{x}$, $U_{y}$, and pressure at $N_{y}=30$ along $x$ direction, at $t=6.0$. The incident shock wave and reflected shock wave are captured by DBM clearly. The simulation results $(\rho,p)_{2}=(1.85129,73,73463)$ at pre-shock wave of first shock wave have a relative error $(-2.1\%,-0.50\%)$ with analytical solutions $(\rho,p)_{theory}=(1.89157,74,10340)$. The analytical solutions are obtained by substituting the coming shock wave $(\rho,p)_{1}=(1.0,1/3.320)$ into the Eqs. (\ref{Eq:DDBM-RH4}) and (\ref{Eq:DDBM-RH5})\cite{2015Xu-PRE}. That two points indicate the ability of capturing two-dimensional shock wave accurately of this two-fluid DBM.
\begin{equation} \label{Eq:DDBM-RH4}
\frac{p_{2}}{p_{1}}=\frac{2\gamma}{\gamma +1}\tt{M}_{1}^{2} \sin^{2}\alpha-\frac{\gamma -1}{\gamma +1}
\tt{,}
\end{equation}
\begin{equation}\label{Eq:DDBM-RH5}
\frac{\rho_{2}}{\rho_{1}}=\frac{(\gamma +1)\tt{M}_{1}^{2} \sin^2\alpha}{(\gamma -1)\tt{M}_{1}^{2} \sin^2\alpha +2}
\tt{,}
\end{equation}

\subsubsection{The TNE behaviours on regular reflection}
\begin{figure*}
\begin{center}
\includegraphics[width=0.8\textwidth]{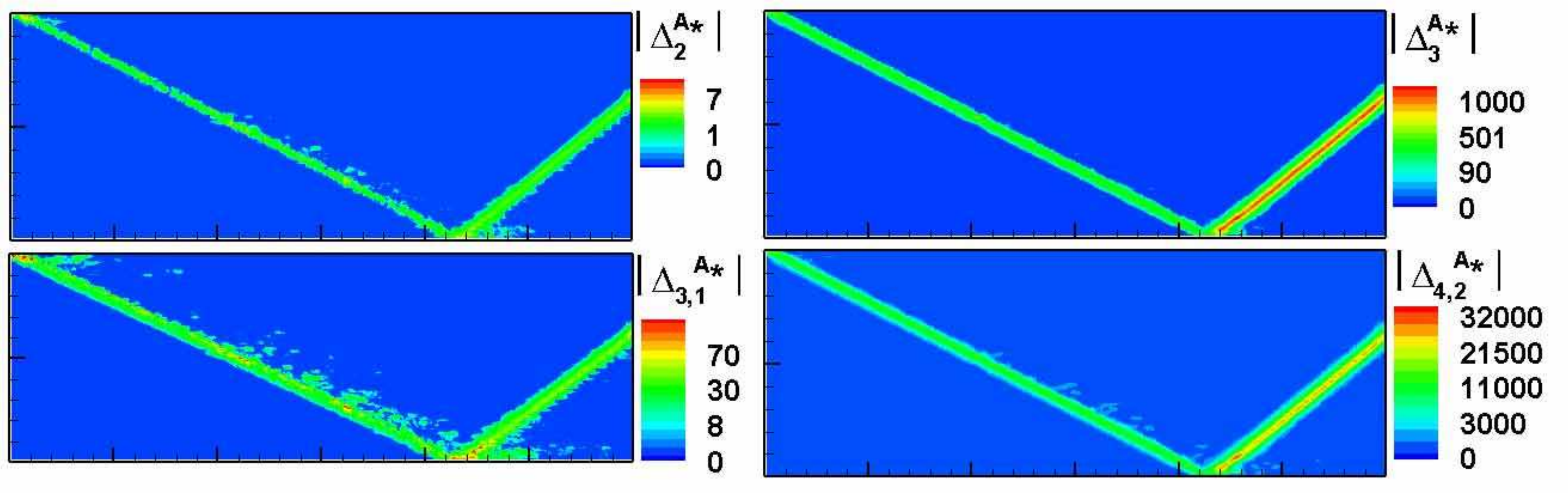}
\end{center}
\caption{$\left|\bm{\Delta}_{2}^{\sigma*}\right|$, $\left|\bm{\Delta}_{3,1}^{\sigma*}\right|$, $\left|\bm{\Delta}_{3}^{\sigma*}\right|$, and $\left|\bm{\Delta}_{4,2}^{\sigma*}\right|$ contours with $\Pr = 1.0$ at $t = 6.0$, respectively. The color from blue to red indicates the increase of values.} \label{fig22}
\end{figure*}

To investigate the TNE behaviors of regular reflection of a shock wave, we give the contours of $\left|\bm{\Delta}_{2}^{A*}\right|$, $\left|\bm{\Delta}_{3,1}^{A*}\right|$, $\left|\bm{\Delta}_{3}^{A*}\right|$, and $\left|\bm{\Delta}_{4,2}^{A*}\right|$ with $\Pr = 1.0$ at $t = 6.0$, respectively. As shown in Fig. \ref{fig22}, the values of four kinds of TNE behaviors are all great than zero around the two shock wave interfaces because of the strong physical quantity gradient, while approaching zero where far away from shock wave interface.

\subsection{Shock wave act on a cylindrical bubble problem}

\begin{figure}
\begin{center}
\includegraphics[width=0.4\textwidth]{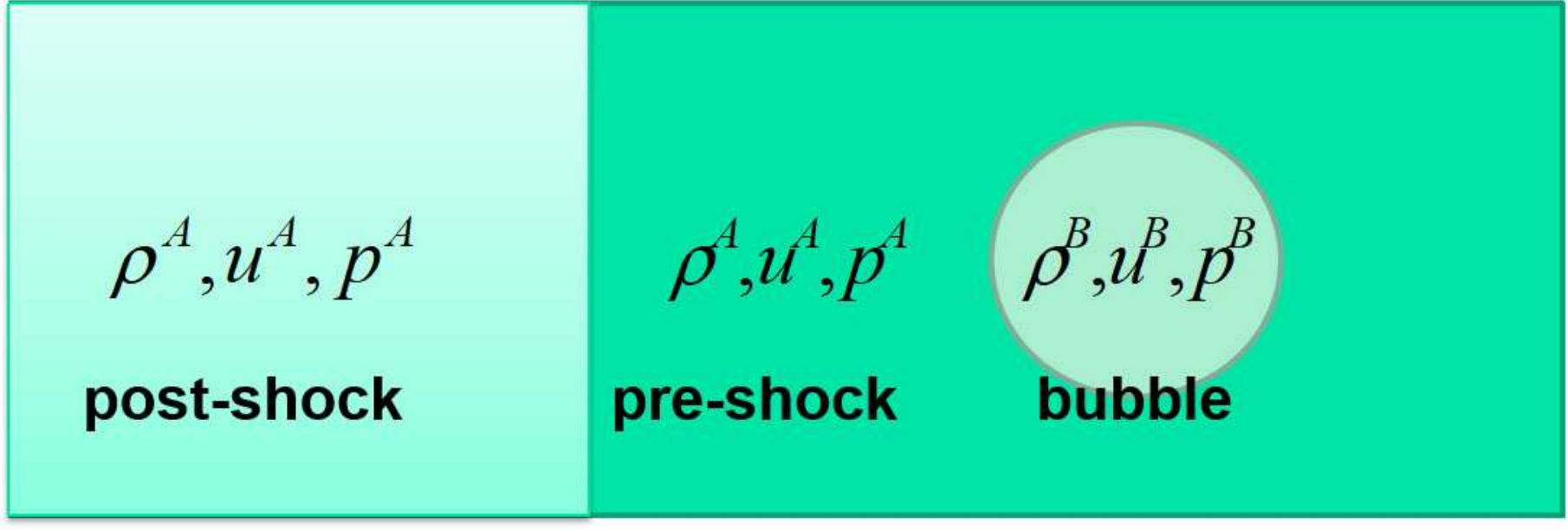}
\end{center}
\caption{Initial field of shock wave act on cylindrical bubble problem. The flow filed is initially divided into three parts: pre-shock, post-shock, and bubble.} \label{fig23}
\end{figure}

The problem of shock wave act on a cylindrical bubble is a classic two-dimensional compressible viscous flow\cite{2010Chen-MRT}. We present this unsteady benchmark problem by our two-fluid model and compare DBM results with other numerical method in previous literatures. In this computational domain with $400\times150$ rectangular grids as shown in Fig. \ref{fig23}, a Mach 1.2 planar shock impinges on a high density cylindrical bubble. The flow filed is initially divided into three parts: pre-shock, post-shock, and bubble. The first two parts have just component A and the third part represent component B, respectively. Initial conditions of pre-shock area, post-shock area and bubble area are as follows:
\[
\left\{
\begin{array}{l}
(\rho,U_{x},U_{y},p)^{A}_{x,y}=(1.0,0.0,0.0,1.0)    \\
(\rho,U_{x},U_{y},p)^{A}_{x,y}=(1.34161,0.361538,0.0,1.51332)\\
(\rho,U_{x},U_{y},p)^{B}_{x,y}=(5.04,0.0,0.0,1.0)
\end{array}
\right.
\]

\begin{figure}
\begin{center}
\includegraphics[width=0.4\textwidth]{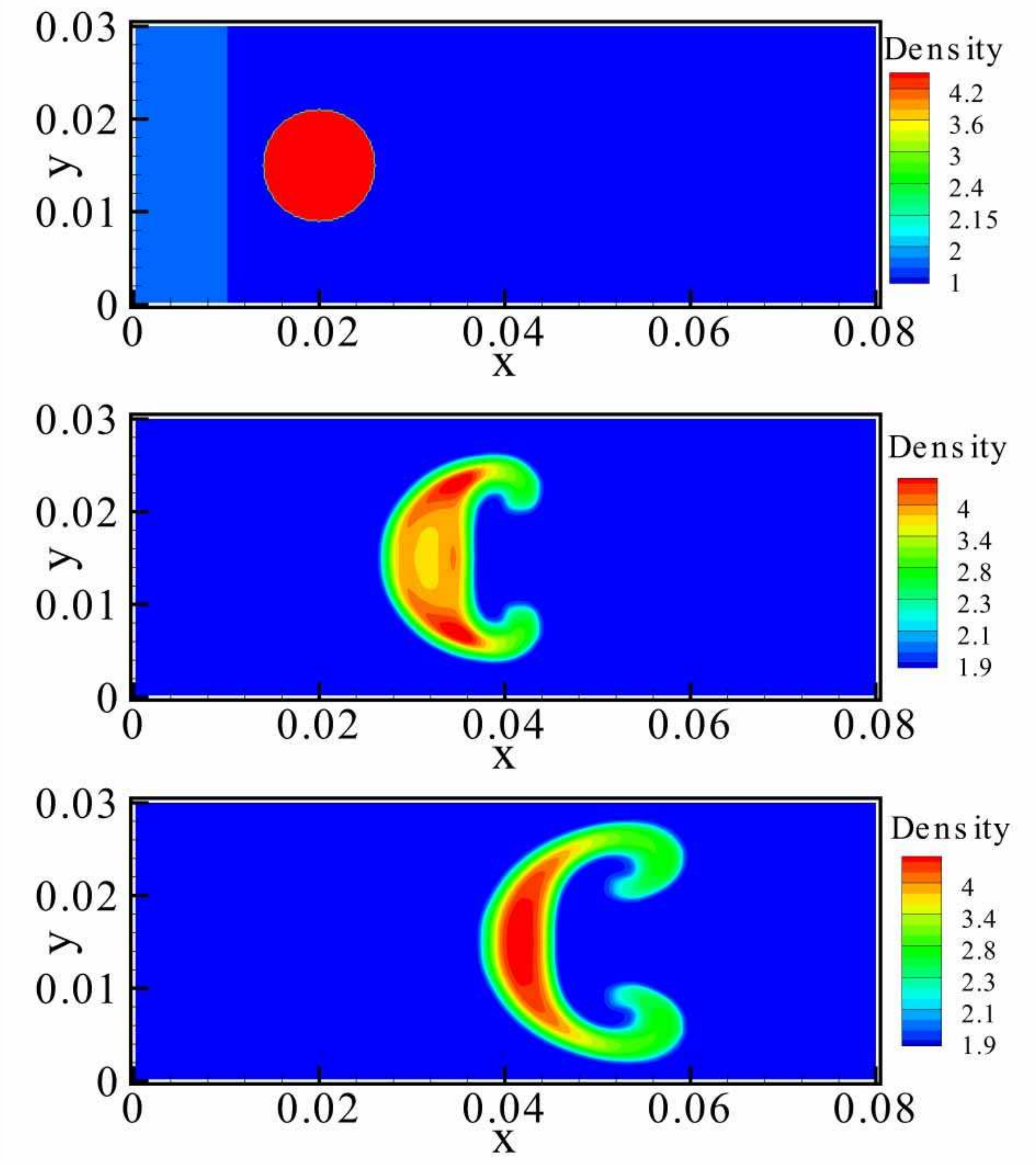}
\end{center}
\caption{Density contours of system in the evolutions at different time $t$=0.0, 0.8, and 1.2, respectively, with $\Pr=1.0$. The color from blue to red indicates the increase of density.} \label{fig24}
\end{figure}

Parameters are as follows: $m^{A}=m^{B}=1.0$, $\Delta t=2\times 10^{-6}$, $\Delta x=\Delta y=2\times 10^{-4}$, $\tau^{A}=5\times 10^{-6}$, $\tau^{B}=8\times 10^{-6}$, $I^{A}=3.0$, $I^{B}=15.0$, $b=0.0$, $c=0.8$, $\eta^{A}=\eta^{B}=10.0$. In this case, inflow and outflow boundary conditions are adopted on the left and right sides of computational domain, and periodic conditionals are imposed on the top and bottom, respectively. From Fig. \ref{fig24} we can see the density contours of physical system on the evolution at three different times, with $\Pr=1.0$. The simulation results are accordant with those by other numerical methods\cite{2010Chen-MRT,2004Shock}.

\section{Conclusions}\label{Conclusions}

 A two-fluid simplified Boltzmann equation based on the ES-BGK model is derived. Then, a two-fluid DBM based on the ES-BGK is formulated for compressible flows. Mathematically, the model is composed of two coupled DBE. Each DBE describes one component of the fluid. Physically, the model is equivalent to a macroscopic fluid model based on Navier-Stokes equations, and supplemented by a coarse-grained model for thermodynamic non-equilibrium behaviors. The model has flexible Prandtl number or specific heat ratio.
 For multiple component mixture, the correspondence between the macroscopic fluid model and the DBM may be several-to-one.

Five types of typical benchmark tests are given to confirm the soundness and accuracy of the model.
Among which, a Sod's shock tube of two components with different particle masses are simulated, which can not achieve in Riemann analytical solution or single-fluid DBM.
 A two-dimensional KHI is simulated, and the Prandtl number effects are investigated. Some hydrodynamic and TNE behaviors of KHI evolution, which are not available in a pure NS model or single-fluid DBM, are presented.
A regular reflection of shock wave is simulated, and the TNE behaviors are studied.
Furthermore, we simulate the problem of a two--dimensional shock wave act on a cylindrical bubble, which shows the ability of our new model to describe the two-component shock problem.

\begin{acknowledgments}
The authors thank Chuandong Lin, Yanbiao Gan, Feng Chen, Ge Zhang, Jiahui Song, Yiming Shan, Cheng Chen, and Xin Lin on helpful discussions on DBM.
This work was supported by the National Natural Science Foundation of China (under Grant No.  11772064),
CAEP Foundation (under Grant No. CX2019033), the Strategic Priority Research Program of Chinese Academy of Sciences (Under Grant No. XDA25051000), the opening project of State Key
Laboratory of Explosion Science and Technology (Beijing Institute of Technology) (under Grant No. KFJJ19-01M), China Postdoctoral Science Foundation (under Grant No. 2019M662521), and Scientific Research Foundation of Zhengzhou university (under Grant No. 32211545 ).
\end{acknowledgments}

\begin{appendix}
\section{Appendix}\label{App}
Performing the operator $\frac{1}{2}\sum_{i}(v_{i\alpha}\cdot v_{i\alpha}+\eta_{i}^{\sigma 2})$ to discrete Boltzmann equation (\ref{Eq:DDBM-N19}), we obtain
\begin{eqnarray*}
\begin{aligned}
\frac{\partial}{\partial t}\rho^{\sigma}E^{\sigma}_{T}&+\frac{\partial}{\partial r_{\alpha}}(\rho^{\sigma}E^{\sigma}_{T}+p^{\sigma})u_{\alpha}=\\
&-\frac{b}{2\tau^{\sigma}}(\Delta_{2,xx}^{\sigma*}+\Delta_{2,yy}^{\sigma*})
\tt{,}
\end{aligned}
\end{eqnarray*}
By submitting Eq. (\ref{Eq:DDBM-NF28}) into the equation, we get
\begin{eqnarray*}
\begin{aligned}
\frac{\partial}{\partial t}\rho^{\sigma}E^{\sigma}_{T}&+\frac{\partial}{\partial r_{\alpha}}(\rho^{\sigma}E^{\sigma}_{T}+p^{\sigma})u_{\alpha}=\\
&-\frac{b}{\tau}
(\frac{I^{\sigma}}{2}\frac{n^{\sigma}T}{m^{\sigma}}-\frac{1}{2}\sum_{i}f_{i}^{\sigma}\eta_{i}^{\sigma})
\tt{.}
\end{aligned}
\end{eqnarray*}
For obtaining a common energy equation, the right side of this equation must be equal to zero namely one of two parameter(extra degree of freedom $I^{\sigma}$ and coefficient $b$) must be zero. Our model recover to a two-fluid DBM based on BGK with extra degree of freedom when $b=0$ whereas to a two-fluid DBM based on ES-BGK with flexible Prandtl number when $I^{\sigma}=0$.
\end{appendix}

\section*{Data Availability}
The data that support the findings of this study are available from the corresponding author upon reasonable request.

\section*{References}

\bibliography{2-fluid-pof}

\end{document}